\documentclass[a4paper,12pt]{article}

\usepackage{bbold}
\usepackage{float}
\usepackage{cancel}
\usepackage{subfig}
\usepackage{amsmath, amssymb, setspace,cite}
\usepackage{slashed}
\usepackage{graphicx}
\usepackage[hidelinks]{hyperref}
\usepackage{color}
\usepackage{soul}
\usepackage{xcolor}
\usepackage{braket}
\usepackage{makecell}
\usepackage{hhline}

\numberwithin{equation}{section}
\vspace{0.6cm}
\date{\today} 

\usepackage{tabularx}
\usepackage{longtable} 
\usepackage{multicol}
\usepackage{multirow}
\usepackage{enumitem}

\usepackage{marginnote}
\begin{document}

\def\thefootnote{\fnsymbol{footnote}}

\begin{center}
\Large\bf\boldmath
\vspace*{1.cm} 
Overview of $B \to K^{(*)}\ell \ell$ Theoretical Calculations and~Uncertainties
\unboldmath
\end{center}
\vspace{0.6cm}

\begin{center}
F.~Mahmoudi$^{1,2,3,}$\footnote{Electronic address: nazila@cern.ch}, 
Y. Monceaux$^{1,}$\footnote{Electronic address: y.monceaux@ip2i.in2p3.fr}\\
\vspace{0.6cm}
{\sl $^1$Universit\'e Claude Bernard Lyon 1, CNRS/IN2P3, \\
Institut de Physique des 2 Infinis de Lyon, UMR 5822, F-69622, Villeurbanne, France}\\[0.4cm]
{\sl $^2$Theoretical Physics Department, CERN, CH-1211 Geneva 23, Switzerland}\\[0.4cm]
{\sl $^3$Institut Universitaire de France (IUF), 75005 Paris, France}
\end{center}

\renewcommand{\thefootnote}{\arabic{footnote}}
\setcounter{footnote}{0}

\vspace{1.cm}
\begin{abstract}
The search for New Physics (NP) beyond the Standard Model (SM) has been a central focus of particle physics, including in the context of $B$-meson decays involving $b \to s \ell \ell$ transitions. These transitions, mediated by flavour-changing neutral currents, are highly sensitive to small NP effects due to their suppression in the SM. While direct searches at colliders have not yet led to NP discoveries, indirect probes through semi-leptonic decays have revealed anomalies in observables such as the branching fraction $\mathcal{B}(B \to K \mu \mu)$ and the angular observable $P_5'(B \to K^* \mu \mu)$. In order to assess the observed tensions, it is essential to ensure an accurate SM prediction. In this review, we examine the theoretical basis of the $B \to K^{(*)} \ell \ell$ decays, addressing in particular key uncertainties arising from local and non-local form factors. We also discuss the impact of QED corrections to the Wilson coefficients, as well as the effect of CKM matrix elements on the predictions and the tension with the experimental measurements. We discuss the most recent results, highlighting ongoing efforts to refine predictions and to constrain potential signs of NP in these critical decay processes.
\end{abstract}

\newpage

\section{Introduction}

The discovery of the Higgs boson at the LHC in 2012~\cite{CMS:2012qbp, ATLAS:2012yve} marked a significant milestone in validating the Standard Model (SM) of particle physics. However, the~SM remains an incomplete theory; thus, the search for New Physics (NP) signals including NP particles has been ongoing at colliders. Although~direct searches have not yet led to discoveries, several hints of potential NP have been observed via indirect searches. 
Semi-leptonic $B$-meson decays via $b \to s \ell^+ \ell^-$ transitions have been prime candidates for probing NP indirectly. These transitions mediated by flavour-changing neutral currents are forbidden at the tree level in the SM and are further suppressed by elements of the Cabibbo--Kobayashi--Maskawa (CKM) matrix, making them particularly sensitive to small NP effects. Numerous deviations from SM predictions, often referred to as anomalies, have been observed in these decays. An~intriguing tension was found in the lepton flavour universality ratios $R_{K^{(*)}}$ of $B \to K^{(*)} \ell^+ \ell^-$ where the ratio is between a final state with a muon pair and an electron pair. However, recent results from LHCb now indicate an SM-like behaviour with no significant deviations observed~\cite{LHCb:2022vje}. Still, deviations remain in other observables such as the branching fraction $\mathcal{B}(B \to K \mu^+ \mu^-)$~\cite{Bell:2014zya, BaBar:2012mrf, LHCb:2014cxe, CMS:2024syx}, and~the angular observable $P_5'(B \to K^* \mu^+ \mu^-)$~\cite{LHCb:2020gog, LHCb:2020lmf, CMS:2017rzx, Belle:2016fev, ATLAS:2018gqc, CMS-PAS-BPH-21-002} at low $q^2$ where $q^2$ denotes the invariant squared mass of the dilepton in the final state. Recent measurements by CMS~\cite{CMS:2024syx, CMS-PAS-BPH-21-002} confirm previous observations by LHCb~\cite{LHCb:2014cxe, LHCb:2020gog, LHCb:2020lmf}, indicating a persistent tension with {the theoretical}~predictions.

However, branching fractions and, to~a lesser degree, the~optimised angular observables $P_i$, are more challenging to predict than the $R_{K^{(*)}}$ ratios and suffer from larger~uncertainties.

In this review, we address the calculation of $B \to K^{(*)} \ell \ell$ observables. In~Section~\ref{sec:theory}, we introduce the notations and key observables. We then turn to the current state of local and non-local form factors in Sections~\ref{sec:form_factors} and~\ref{sec:non_local}, which constitute the primary sources of uncertainty in these calculations. In~Section~\ref{sec:accuracy}, we address the impact of QED corrections to Wilson coefficients as well as the effect of CKM matrix elements, which become increasingly significant as precision increases. Section~\ref{sec:impact} presents a quantitative assessment of the implications of different theoretical predictions. Finally, Section~\ref{sec:Conclusion} provides our~conclusions.

\section{Theoretical~Framework}\label{sec:theory}

$B \to K^{(*)} \ell^+  \ell^-$ decays are well described in the Weak Effective Theory. In~this formalism, the~transition $b \to s \ell^+ \ell^-$ is described by an effective Hamiltonian, where degrees of freedom above the electroweak scale have been integrated out~\cite{Buchalla:1995vs, Bobeth:1999mk, Bobeth:2001jm}:

\begin{equation} \label{eq:Heff}
\small
    \mathcal{H}_{\text{eff}} = -\frac{4G_F}{\sqrt{2}} \left\{\lambda_t \bigg(\sum_{i=1}^{i=2} C_i \mathcal{O}_i^c + \sum_{i=3}^{i=6} C_i \mathcal{O}_i + \sum_{i=7, 8, 9, 10} (C_i \mathcal{O}_i + C_i^{'} \mathcal{O}_i^{'})\bigg) + \lambda_u  \bigg(\sum_{i=1}^{i=2} C_i (\mathcal{O}_i^c - \mathcal{O}_i^u) \bigg) \right\} + \text{h.c}.
\end{equation}

The CKM factor $\lambda_j$ denotes $\lambda_j = V_{jb} V_{js}^*$ and $G_F$ is the Fermi coupling constant. The~local operators $\mathcal{O}_i$ and their associated Wilson coefficients $C_i$ are given in the standard basis introduced in~\cite{Chetyrkin:1997gb} by

\begin{equation}\label{eq:local_operators}
    \begin{split}
        \begin{aligned}
            \mathcal{O}_1^q = & (\bar{s}_L \gamma_{\mu}T^a q_L)
                                (\bar{q}_L  \gamma^{\mu}T^a b_L )\, , \\
            \mathcal{O}_3 = & (\bar{s}_L  \gamma_{\mu} b_L )
                              \sum_{q'} (\bar{{q'}}   \gamma^{\mu} {q'} )\, , \\
            \mathcal{O}_5 = & (\bar{s}_L  \gamma_{\mu_1}\gamma_{\mu_2}\gamma_{\mu_3} b_L )
                              \sum_{q'} (\bar{{q'}}   \gamma^{\mu_1}\gamma^{\mu_2}\gamma^{\mu_3} {q'} )\, , \\
            \mathcal{O}_7 = & \frac{e}{16 \pi^2}m_b(\bar{s} \sigma^{\mu \nu}P_R b )F_{\mu \nu}\, , \\
            \mathcal{O}_9 = & \frac{e^2}{16 \pi^2}(\bar{s}  \gamma_{\mu} P_L b) (\bar{l}\gamma^\mu l)\, ,
        \end{aligned}
        \qquad 
        \begin{aligned}
            \mathcal{O}_2^q = & (\bar{s}_L \gamma_{\mu} q_L )
                                (\bar{q}_L  \gamma^{\mu} b_L )\, , \\
            \mathcal{O}_4 = & (\bar{s}_L  \gamma_{\mu}T^a b_L )
                              \sum_{q'} (\bar{{q'}}   \gamma^{\mu}T^a {q'} )\, , \\
            \mathcal{O}_6 = & (\bar{s}_L  \gamma_{\mu_1}\gamma_{\mu_2}\gamma_{\mu_3} T^ab_L )
                              \sum_{q'} (\bar{{q'}} \gamma^{\mu_1}\gamma^{\mu_2}\gamma^{\mu_3} T^a {q'} )\, , \\
            \mathcal{O}_8 = & \frac{g_s}{16 \pi^2}m_b(\bar{s} \sigma^{\mu \nu}P_R T^a b )G^a_{\mu \nu}\, , \\
            \mathcal{O}_{10} = & \frac{e^2}{16 \pi^2}(\bar{s} \gamma_{\mu} P_L b) (\bar{l}\gamma^\mu \gamma_5 l)\, ,
        \end{aligned}
    \end{split}
\end{equation}
where $g_s$ is the strong coupling constant; the~quark flavour $q=u, c$; the~lepton flavour ${l=e, \mu, \tau}$; and $m_b$ is the running $b$-quark mass in the $\overline{\text{MS}}$ scheme. We use the conventions ${P_{L, R}~=~(1 \mp \gamma_5)/2}$ and $\sigma_{\mu \nu} = \frac{i}{2}\left[\gamma_\mu, \gamma_\nu \right]$. The~primed local operators $\mathcal{O}_i$ are obtained by performing the exchange $P_L \leftrightarrow P_R$.  

The term proportional to $\lambda_u$ in the definition \eqref{eq:Heff} is often neglected as it is strongly CKM-suppressed with respect to the term proportional to $\lambda_t$. However, it can be relevant for observables that are specifically sensitive to complex phases of decay amplitudes. We discard it in the~following. 

The decay amplitude reads as follows:
\begin{equation}\label{eq:general_amplitude}
    A(B \to K^{(*)} \ell^+ \ell^-) = - \bra{K^{(*)}(k) \ell^+ \ell^-} \mathcal{H}_{\text{eff}}\ket{B(p_B = k+q)}\, ,
\end{equation}
which leads to the expression~\cite{Khodjamirian:2010vf, Gubernari:2020eft, Gubernari:2022hxn}
\begin{align} \label{eq:detailed_amplitude}
    A(B \to K^{(*)} \ell^+ \ell^-)
    \equiv
        \frac{G_F\, \alpha\, V_{tb}^{} V_{ts}^*}{\sqrt{2} \pi}
       & \bigg\{ (C_9 \,L^\mu_{V} + C_{10} \,L^\mu_{A})\ \! \mathcal{F}_\mu^{B \to K^{(*)}}
        \! \nonumber \\
        &-  \frac{L^\mu_{V}}{q^2} \Big[  2 i m_b C_7\,\mathcal{F}_{T, \mu}^{B \to K^{(*)}}
        \! + 16\pi^2 \mathcal{H}_\mu^{B \to K^{(*)}} \Big]   \bigg\}
        \, .
\end{align}

Here, $\alpha$ denotes the electromagnetic coupling constant,~$L^\mu_{V, A}$ are leptonic currents, and $\mathcal{F}_{(T),\mu}$ and $\mathcal{H}_\mu$ are, respectively, local and non-local hadronic matrix elements. They are given by
\begin{align}
    L^\mu_V & \equiv \bar{u}_\ell(q_1)\gamma^\mu v_\ell(q_2)\,, \\
    L^\mu_A & \equiv \bar{u}_\ell(q_1)\gamma^\mu \gamma_5 v_\ell(q_2)\,, \\
    \mathcal{F}_\mu^{B \to K^{(*)}} & \equiv \bra{\bar{K}^{(*)}(k)}\bar{s}\gamma_\mu P_L b \ket{\bar{B}(p_B = k+q)}\, , \\
    \mathcal{F}_{T, \mu}^{B \to K^{(*)}} & \equiv \bra{\bar{K}^{(*)}(k)}\bar{s}\sigma_{\mu \nu} q^\nu P_L b \ket{\bar{B}(p_B = k+q)}\, , \\
    \mathcal{H}_\mu^{B \to K^{(*)}} & \equiv \sum_{q'} \mathcal{H}_{q', \mu}^{B \to K^{(*)}}\, . \label{eq:non_local_sum}
\end{align}

In the last term $\mathcal{H}_\mu^{B \to K^{(*)}}$, the sum runs over the accessible quark flavours (at the typical scale $\mu_b = m_b$) $q' = u, d, s, c, b$. 
\newpage
\noindent For~a given quark flavour,
\begin{align}\label{eq:non_local}
    \mathcal{H}_{q', \mu}^{B \to K^{(*)}} (q,k) \equiv & i Q_{q'} \int  d^4x e^{iq \cdot x} \nonumber\\ 
    & \times \bra{\bar{K}^{(*)}(k)} T \Big\{\bar{q'}\gamma_\mu q'(x), \Big(\sum_{i=1}^{i=2} C_i \mathcal{O}_i^c + \sum_{i=3}^{i=6} C_i \mathcal{O}_i + C_8 \mathcal{O}_8 \Big)(0)\Big\}\ket{\bar{B}(k+q)},
\end{align}
where $Q_{q'}$ denotes the electric charge of the quark $q'$. The~local operators $\mathcal{O}_1^c$ and $\mathcal{O}_2^c$ have been singled out in \eqref{eq:Heff} and \eqref{eq:non_local}, as~they numerically contribute more than the other hadronic operators $\mathcal{O}_{3-6}$. 

The expression \eqref{eq:detailed_amplitude} is obtained at leading order in QED. The~QED effects have been discussed in~\cite{Isidori:2020acz, Isidori:2022bzw, Choudhury:2023uhw} where the $\mathcal{O}(\alpha)$ corrections have been computed. For~the partial width, the~reduction is estimated to be up to $10 \%$ at high $q^2$ for muons in the final state, while it is even larger when considering electrons in the final state. Such effects are accounted for on the experimental side with software like \texttt{PHOTOS}~\cite{Golonka:2005pn} before comparison with the theoretical~predictions.

In the following, we describe briefly the calculation of $B \to K \ell^+ \ell^-$ and $B \to K^* \ell^+ \ell^-$ observables. For~detailed descriptions, we refer the reader to the \texttt{SuperIso} manual~\cite{Mahmoudi:2008tp}.

\subsection[B -> K l l]{$B \to K \ell^+ \ell^-$}
 \label{sec:B_to_K_l_l}
The full differential distribution of the $B \to K \ell^+ \ell^-$ decay can be expressed in the SM as~\cite{Becirevic:2012fy, Bobeth:2007dw}
\begin{equation} \label{eq:K_differential}
    \frac{d^2 \Gamma (\bar{B} \to \bar{K} \ell^+ \ell^-)}{dq^2 d\cos{\theta}} = a_\ell(q^2) + c_\ell(q^2) \cos{\theta}^2\, ,
\end{equation}
where $\theta$ is defined as the angle between the directions of the lepton $\ell^-$ and the $\bar{B}$-meson in the rest frame of the lepton pair. The~boundaries of the phase space are given by
\begin{equation}
    4m_\ell^2 \leq q^2 \leq (m_B - m_K)^2, \hspace{1cm} -1 \leq \cos{\theta} \leq 1\, .
\end{equation}

The functions $a_\ell$ and $c_\ell$ in Equation~\eqref{eq:K_differential}
are defined in the SM as follows:
\begin{align}\label{coeff_BKell}
a_\ell  (q^2) = &\  {\mathcal{C}}(q^2)\Big[  q^2 \lvert F_P(q^2) \rvert^2   +
 \frac{\lambda(m_B^2, m_K^2, q^2)   }{4}  \left( \lvert F_A (q^2)\rvert^2 +  \lvert F_V(q^2) \rvert^2 \right)    \nonumber  \\
 & \hspace{1.25cm}+ 4 m_{\ell}^2 m_B^2 \lvert F_A(q^2) \rvert^2+2m_{\ell}  \left( m_B^2 -m_K^2 +q^2\right) \text{Re}\left( F_P(q^2) F_A^{\ast}(q^2) \right)\Big]\, ,   \\
& \nonumber \\
c_\ell  (q^2) = &\  {\mathcal{C}}(q^2)\Big[ - \frac{\lambda(m_B^2, m_K^2, q^2)  }{4} \beta_{\ell}^2(q^2) \left( \lvert F_A(q^2) \rvert^2 +  \lvert F_V(q^2) \rvert^2 \right) \Big]\, ,
\end{align}
with the prefactor
\begin{equation}
    \mathcal{C}(q^2) \equiv \frac{G_F^2 \alpha^2 \lvert V_{tb}V_{ts}^* \rvert^2}{512 \pi^5 m_B^3} \beta_\ell(q^2)\sqrt{\lambda(m_B^2, m_K^2, q^2)}\, ,
\end{equation}
where $\beta_\ell(q^2) \equiv \sqrt{1-4\frac{m_\ell^2}{q^2}}$ for $\ell = e, \mu, \tau$, and~$\lambda$ is the Källén function:
\begin{equation}
    \lambda(x, y, z) = x^2 + y^2 + z^2 - 2(x y + y z + x z)\, .
\end{equation}

In the above equations, $F_V, F_A$, and $F_P$ can be written in the SM as~\cite{Becirevic:2012fy}
\begin{align}\label{eq:Fi_def}
    F_V(q^2) = &\ (C_9 + C_9^{'})f_+(q^2) + \frac{2 m_b}{m_B + m_K}(C_7^{\text{eff}} + C_7^{'})f_T(q^2) + \delta F_V\, , \\
    F_A(q^2) = &\ (C_{10} + C_{10}^{'}) f_+(q^2)\, , \\
    F_P(q^2) = &\ -m_l (C_{10} + C_{10}^{'}) \Big[f_+(q^2) - \frac{m_B^2-m_K^2}{q^2}(f_0(q^2) - f_+(q^2)) \Big]\, ,
\end{align}
where $C_7^{\text{eff}}$ is defined in Equation~(\ref{eq:C7eff}). The~local form factors $f_+, f_0$, and $f_T$ are defined in Appendix \ref{appendix:FFdef} and their calculations are discussed in Section~\ref{sec:form_factors}. The~term $\delta F_V$ corresponds to non-local contributions and is addressed in more detail in Section~\ref{sec:non_local}. Appendix~\ref{appendix:alternate_basis} introduces the alternative form factor basis proposed in~\cite{Gubernari:2022hxn}, establishing the correspondence between their transversity amplitudes and the $F_i$ functions, as~well as the correspondence of the non-local term.  \\[1\baselineskip]
\textbf{Observables} \\

It is customary to introduce the $q^2$-integrated coefficients~\cite{Bobeth:2007dw, Becirevic:2012fy}
\begin{equation}
    A_\ell = \int_{q^2_{\text{min}}}^{q^2_{\text{max}}} dq^2 a_\ell(q^2)\,, \hspace{0.75cm} C_\ell = \int_{q^2_{\text{min}}}^{q^2_{\text{max}}} dq^2 c_\ell(q^2)\, ,
\end{equation}
to express the observables. The~decay rate can then be written as
\begin{equation}
    \Gamma(B \to K \ell^+ \ell^-) = 2 \Big(A_\ell + \frac{1}{3} C_\ell \Big)\, ,
\end{equation}
and the flat-term is
\begin{equation}
    F_H^\ell = \frac{2}{\Gamma_\ell}(A_\ell + C_\ell)\, .
\end{equation}

In the SM, the forward--backward asymmetry is null and the flat term is proportional to $m_\ell$. However, they can receive sizeable NP contributions, which make them relevant~observables.

\subsection[B K* l l]{$B \to K^* \ell^+ \ell^-$}\label{sec:B_to_Kstar_l_l}
For $\bar{B} \to\bar{ K}^* \ell^+ \ell^-$, the~process that is measured is $\bar{B} \to\bar{ K}^* (\to K \pi) \ell^+ \ell^-$. The~subsequent decay $K^* \to K \pi$ can be described with the effective Hamiltonian~\cite{Kim:2000dq}:
\begin{equation}
    \mathcal{H_{\text{eff}}} = g_{K^*K\pi}(p_K - p\pi) \cdot \varepsilon_{K^*}\, ,
\end{equation}
where $g_{K^*K\pi}$ is the coupling constant and $\varepsilon_{K^*}$ is the polarisation of the $K^*$ meson. It is convenient to consider the $K^*$-meson on the mass-shell when using a narrow-width approximation~\cite{Kim:2000dq, Altmannshofer:2008dz, Kruger:1999xa} and to replace the squared $K^*$ propagator by 

\begin{equation}
    \frac{1}{(p_{K^*}^2 - m_{K^*}^2)^2 +(m_{K^*}\Gamma_{K^*})^2} \,\underset{\Gamma_{K^*} \ll m_{K^*}}{\longrightarrow} \, \frac{\pi}{m_{K^*}\Gamma_{K^*}}\delta(p_{K^*}^2 - m_{K^*}^2)\, .
\end{equation}

Since the width of the $K^*$ meson can be written as
\begin{equation}
    \Gamma_{K^*} = \frac{g_{K^* K \pi}^2}{48 \pi} m_{K^*} \beta^3,
\end{equation}
where $\beta$ is related to the Källén function as
\begin{equation}
    \beta = \frac{1}{m_{K^*}^2}\lambda(m_{K^*}^2, m_K^2, m_\pi^2)^{1/2}\, ,
\end{equation}
the final result in this narrow-width limit is independent of the coupling $g_{K^* K \pi}$ which cancels out. The~impact of the finite width of the $K^*$-meson is addressed in Section~\ref{sec:form_factors}. 

In the narrow-width approximation, summing over the lepton spins, the~full differential distribution can be written as~\cite{Egede:2008uy, Bobeth:2008ij, Egede:2010zc, Altmannshofer:2008dz, Kim:2000dq}
\begin{equation}
    \frac{d^4 \Gamma}{dq^2 d\cos{\theta_\ell}d \cos{\theta_{K^*}}d\phi} = \frac{9}{32 \pi} J(q^2, \theta_\ell, \theta_{K^*}, \phi)\, .
\end{equation}

We work in the convention where $\theta_\ell$ is defined as the angle between the directions of the lepton $\ell^-$ and the $\bar{B}$-meson in the rest frame of the lepton pair, $\theta_{K^*}$ is the angle between the directions of the $K$-meson and the $\bar{B}$-meson in the $K \pi$ rest frame, and $\phi$ is the angle between the normals of the plane of $K \pi$ and the plane of the lepton pair. In~\cite{Altmannshofer:2008dz, Egede:2008uy}, the~convention for $\theta_\ell$ is slightly different. More details about the angles and conventions are given in Appendix \ref{appendix:angular}. 

The boundaries of the phase space are

\begin{equation}
    4 m_\ell^2 \leq q^2 \leq (m_B - m_{K^*})^2\,, \hspace{0.5cm} -1 \leq \cos{\theta_\ell} \leq 1\,, \hspace{0.5cm} -1 \leq \cos{\theta_{K^*}} \leq 1\,, \hspace{0.5cm}  0 \leq \phi \leq 2 \pi \,.
\end{equation}

The explicit expression of $J$ is
\begin{align}\label{eq:explicit_J}
    J(q^2,\theta_k,\theta_l,\phi) =  &  J_1^c\cos^2{\theta_k} 
     + J_1^s \sin^2{\theta_k}
     + (J_2^c\cos^2{\theta_k}  + J_2^s\sin^2{\theta_k} ) \cos{2\theta_l}
     + J_3\sin^2{\theta_k}\sin^2{\theta_l} \cos{2\phi} \nonumber  \\
     &+ J_4\sin{2\theta_k}\sin{2\theta_l}\cos{\phi} 
     + J_5\sin{2\theta_k}\sin{\theta_l}\cos{\phi} 
     + J_6\sin^2{\theta_k}\cos{\theta_l} \nonumber\\ 
    & + J_7\sin{2\theta_k}\sin{\theta_l}\sin{\phi} \
     + J_8 \sin{2\theta_k}\sin{2\theta_l}\sin{\phi} 
     + J_9 \sin^2{\theta_k}\sin^2{\theta_l}\sin{2\phi}\, . 
\end{align}

The angular coefficients $J_i^a$ with $i=1, \ldots, 9$ and $a = s, c$ can be expressed with the transversity amplitudes $A_0, A_\parallel, A_\perp$:
\begin{align}
    & J_1^c = |A_0^L|^2 + |A_0^R|^2 + \frac{4m_l^2}{q^2} \big(|A_{t}|^2 + 2\operatorname{Re}(A_0^LA_0^{L*})\big)\, , &
    J_2^c = -\beta_l^2 (|A_0^L|^2 + |A_0^R|^2)\, , \notag \\
    & J_1^s = \frac{2+\beta_l^2}{4} \big(|A_{\perp}^L|^2 + |A_{\parallel}^L|^2 + (L \leftrightarrow R) \big) &
    J_2^s = \frac{\beta_l^2}{4} \big(|A_{\parallel}^L|^2 + |A_{\perp}^L|^2 + (L \leftrightarrow R)\big)\, , \notag \\
    & \quad \quad + \frac{4 m_l^2}{q^2} \big(\operatorname{Re}(A_{\parallel}^L A_{\parallel}^{R*}) + \operatorname{Re}(A_{\perp}^LA_{\perp}^{R*}) \big)\,, &
    J_3 = \frac{\beta_l^2}{2} \big(|A_{\perp}^L|^2 - |A_{\parallel}^L|^2 + (L \leftrightarrow R) \big)\,, \notag \\
    & J_4 = \frac{\beta_l^2}{\sqrt{2}} \big(\operatorname{Re}(A_0^LA_{\parallel}^{L*}) + (L \leftrightarrow R) \big)\,, &
    J_5 = \sqrt{2} \beta_l \big(\operatorname{Re}(A_0^LA_{\perp}^{L*}) - (L \leftrightarrow R) \big)\,, \notag \\
    & J_6 = 2 \beta_l \big(\operatorname{Re}(A_{\parallel}^LA_{\perp}^{L*}) - (L \leftrightarrow R) \big)\,, &
    J_7 = \sqrt{2} \beta_l \big(\operatorname{Im}(A_0^LA_{\parallel}^{L*}) - (L \leftrightarrow R) \big)\,, \notag \\
    & J_8 = \frac{\beta_l^2}{\sqrt{2}} \big(\operatorname{Im}(A_0^LA_{\perp}^{L*}) + (L \leftrightarrow R) \big)\,, &
    J_9 = \beta_l^2 \big(\operatorname{Im}(A_{\parallel}^{L*} A_{\perp}^L) + (L \leftrightarrow R) \big)\,, 
\end{align}

where, again, $\beta_\ell(q^2) = \sqrt{1-4\frac{m_\ell^2}{q^2}}$. 

The transversity amplitudes can be written as follows:
\small
\begin{align}
    A_0^{L, R} &= -\frac{N}{2m_{K^*}\sqrt{q^2}} \bigg[\bigg((C_9-C_9^{'})\mp (C_{10}-C_{10}^{'})\bigg)[(m_B^2-m_{K^*}^2-q^2)(m_B+m_{K^*})A_1  \nonumber\\
    & - \frac{\lambda(m_B^2, m_{K^*}^2, q^2)}{m_B+m_{K^*}}A_2] + 2m_b(C_{7}^{\text{eff}}-C_{7}^{'})[(3m_{K^*}^2+m_B^2 - q^2 )T_2 - \frac{\lambda(m_B^2, m_{K^*}^2, q^2)}{m_B^2-m_{K^*}^2}T_3]\bigg] + \delta A_0^{L, R}\,, \nonumber\\
    A_\perp^{L, R} &= N \sqrt{2\lambda(m_B^2, m_{K^*}^2, q^2)}\bigg[\bigg((C_9+C_9^{'})\mp (C_{10}+C_{10}^{'})\bigg)\frac{V}{m_B+m_{K^*}}+2m_b\frac{C_{7}^{\text{eff}}+C_{7}^{'}}{q^2}T_1\bigg] + \delta A_\perp^{L, R}\,,\nonumber\\
    A_\parallel^{L, R} &= -N \sqrt{2}\bigg[\bigg((C_9-C_9^{'})\mp (C_{10}-C_{10}^{'})\bigg)(m_B+m_{K^*})A_1+2m_b\frac{C_{7}^{\text{eff}}-C_{7}'}{q^2}(m_B^2-m_{K^*}^2)T_2\bigg]+ \delta A_\parallel^{L, R}\,,\nonumber\\
    A_{t} & = 2N (C_{10} - C_{10}^{'})\frac{\sqrt{\lambda(m_B^2, m_{K^*}^2, q^2)}}{\sqrt{q^2}}A_0\,,
\end{align}\normalsize
with the prefactor
\begin{equation*}
    N = V_{tb}V^*_{ts}\bigg[ \frac{\alpha^2 G_F^2}{3\times 2^{10}\pi^5 m_B^3}q^2\beta_l \sqrt{\lambda(m_B^2, m_{K^*}^2, q^2)} \bigg]^{1/2}\,.
\end{equation*}

The local form factors $V, A_0, A_1, A_2, T_1, T_2$, and $T_3$ are defined in Appendix \ref{appendix:FFdef} and their calculations are discussed in Section~\ref{sec:form_factors}. The~terms $\delta A_{i}$ correspond to non-local contributions, which are addressed in more detail in Section~\ref{sec:non_local}. Appendix~\ref{appendix:alternate_basis} introduces the alternative form factor basis suggested in~\cite{Gubernari:2022hxn}, and~establishes the correspondence between their transversity amplitudes and the ones introduced in this section, as~well as the correspondence of the non-local~terms.

For the CP-conjugated decay $B \to K^* \ell^- \ell^+$, the~full differential distribution can be written as~\cite{Bobeth:2008ij, Gratrex:2015hna}
\begin{equation}
     \frac{d^4 \bar{\Gamma}}{dq^2 d\cos{\theta_\ell}d \cos{\theta_{K^*}}d\phi} = \frac{9}{32 \pi} \bar{J}(q^2, \theta_\ell, \theta_{K^*}, \phi)\, . 
\end{equation}

The explicit expression of $\bar{J}(q^2, \theta_\ell, \theta_{K^*}, \phi)$ can be derived from that of $J(q^2, \theta_\ell, \theta_{K^*}, \phi)$ in Equation~\eqref{eq:explicit_J} while performing the following replacements:
\begin{equation}
    J_{1, 2, 3, 4, 7}^{(a)} \to \bar{J}_{1, 2, 3, 4, 7}^{(a)}\,, \hspace{1cm} J_{5, 6, 8, 9} \to - \bar{J}_{5, 6, 8, 9}\,,
\end{equation}
where $\bar{J}_i^{(a)}$ is obtained by conjugating all weak phases in $J_i^{(a)}$. \\
The relative signs when going from $J_i^{(a)}$ to $\bar{J}_i^{(a)}$ can be understood from transforming the angles as $(\theta_l, \theta_{K^*}, \phi) \to (\pi - \theta_l, \pi - \theta_{K^*}, 2\pi - \phi)$, which is the usual convention.
\\[1\baselineskip] 
\textbf{Observables} \\ 

We introduce below some of the key observables for the $\bar{B} \to \bar{K}^* \ell^+ \ell^-$ decay. The~dilepton-invariant mass spectrum is obtained by integrating the full differential distribution over all three angles~\cite{Bobeth:2008ij}:
\begin{equation}
    \frac{d \Gamma}{dq^2} = \frac{3}{4}\Big(J_1 - \frac{J_2}{3} \Big)\,,
\end{equation}
where, for convenience,
\begin{equation}
    J_{1, 2} \equiv 2 J_{1, 2}^s + J_{1, 2}^c\,.
\end{equation}

The normalised forward--backward asymmetry is defined as~\cite{Matias:2012xw}
\begin{align}
    A_{FB}(q^2) &\equiv \Bigg[\int_{-1}^0 - \int_0^1 \Bigg] d \cos{\theta_l} \frac{d^2 \Gamma}{dq^2 d \cos{\theta_l}} \bigg/ \frac{d\Gamma}{dq^2} \nonumber \\
    & = -\frac{3}{8} \frac{2 J_6}{d\Gamma / dq^2}\,.
\end{align}

In~\cite{Bobeth:2008ij, Beaujean:2012uj, Feldmann:2002iw, Bobeth:2010wg, Altmannshofer:2008dz}, the~global sign is different in the definition of the~forward--backward-asymmetry. 

The $K^*$-meson polarisation fractions $F_L$ and $F_T$ are given by~\cite{Matias:2012xw, Beaujean:2012uj, Egede:2008uy, Bobeth:2010wg}
\begin{equation}
    F_L(q^2) = \frac{3 J_1^c - J_2^c}{4 d\Gamma / dq^2}\,, \hspace{1.cm} F_T(q^2) = \frac{4 J_2^s}{d\Gamma / dq^2}\,,
\end{equation}
and the $K^*$-meson polarisation parameter reads as follows~\cite{Egede:2008uy}:
\begin{equation}
    \alpha_{K^*}(q^2) = 2 \frac{F_L}{F_T} - 1\,.
\end{equation}

A set of theoretically clean angular observables has been introduced in~{\cite{Egede:2008uy,Matias:2012xw}}, designed to be less sensitive to form factors. Primed angular observables were later introduced in~\cite{Descotes-Genon:2012isb}. They are defined in the SM as
\begin{align}
    P_1(q^2) &= \frac{J_3}{2 J_2^s}\,, &
    P_2(q^2) &= \beta_\ell \frac{J_6}{8 J_2^s}\,, \notag \\
    P_3(q^2) &= -\frac{J_9}{4 J_2^s}\,, &
    P_4(q^2) &= \frac{\sqrt{2} J_4}{\sqrt{-J_2^c (2 J_2^s - J_3)}}\,, \notag \\
    P_5(q^2) &= \frac{\beta_\ell J_5}{\sqrt{-2 J_2^c (2 J_2^s + J_3)}}\,, &
    P_6(q^2) &= -\frac{\beta_\ell J_7}{\sqrt{-2 J_2^c (2 J_2^s - J_3)}}\,, \notag \\
    P_4'(q^2) &= \frac{J_4}{\sqrt{-J_2^c J_2^s}}\,, &
    P_5'(q^2) &= \frac{J_5}{2\sqrt{-J_2^c J_2^s}}\,, \notag \\
    P_6'(q^2) &= -\frac{J_7}{2\sqrt{-J_2^c J_2^s}}\,. &
\end{align}

For all of the observables given in this section, one can define the CP-average quantities, which are often the ones measured~experimentally. 

\section{Local Form~Factors}\label{sec:form_factors}
Local form factors can be computed in lattice QCD or with QCD sum rules on the light-cone. Lattice QCD determinations, based on first principles, are typically more accurate and reliable but are mostly limited to the low-recoil region (high $q^2$). Light-Cone Sum Rules (LCSRs) can bridge that gap, as~they allow for determinations in the low-$q^2$ region, although~they suffer from systematic uncertainties that are challenging to~evaluate.

We present in the following predictions the full set of local form factors. {Instead, one can} introduce a reduced set of form factors (soft form factors) in the heavy quark limit within the large recoil region. Under~this approximation, only two independent form factors remain for $B \to K^*$ and only one for $B \to K$~\cite{Beneke:2000wa}. This significantly reduces the uncertainties related to the form factors. {Nonetheless, considering the correlations among the full form factors, a similar cancellation of the uncertainties can be obtained.}

\subsection{Lattice~QCD}
\unskip
\subsubsection[B -> K]{$B \to K$}
The latest lattice results for the hadronic $B \to K$ form factors are given in~\cite{Bailey:2015dka} for the FNAL/MILC collaboration, and~in~\cite{Parrott:2022rgu} for the HPQCD collaboration.
The FNAL/MILC results are obtained directly for the range $q^2 \geq 17$ GeV$^2$, and~expanded to the whole $q^2$ range using the Boyd, Grinstein, and~Lebed (BGL) parameterisation~\cite{Boyd:1994tt}. The~total errors, which encompass both the statistical and the systematic uncertainties, are below 4$\%$ at high $q^2$. For~lower $q^2$, the~errors are of the order of 10$\%$ for $f_+$ and around 30$\%$ for $f_T$. 

The recent determination by HPQCD~\cite{Parrott:2022rgu} supersedes their previous results~\cite{Bouchard:2013eph} and covers the entire $q^2$ range, sending the lattice spacing $a \to 0$ and simultaneously fitting them using the Bourreley--Caprini--Lellouch (BCL) parameterisation~\cite{Bourrely:2008za}. This determination is highly precise, with~quoted errors of less than $4 \%$ for $f_+$ and $f_0$ and~less than $7 \%$ for $f_T$. Their results at $q^2=0$ GeV$^2$ read as follows:
\begin{align}
    f_+(q^2 = 0) = f_0(q^2 = 0) &= 0.332 \pm 0.012\,,\\
    f_T(q^2 = 0, \mu=4.8\text{ GeV}) &= 0.332 \pm 0.024\,.
\end{align}

Confirmation of this new determination using the new approach is still awaited from other lattice~collaborations. 

The Flavour Lattice Averaging Group~\cite{FlavourLatticeAveragingGroupFLAG:2021npn} performed an average of the results from~\cite{Bailey:2015dka} and~\cite{Bouchard:2013eph}, which has not yet been updated to incorporate the latest HPQCD results.
\subsubsection[B -> K*]{$B \to K^*$}
Lattice QCD results for the $B \to K^*$ local form factors are given in~\cite{Horgan:2013hoa} and~updated in~\cite{Horgan:2015vla}. These results are obtained directly at low recoil and~extrapolated using their parameterisation. 
At $q^2 = 0$ GeV$^2$, the~results read as follows:
\begin{align}
    V(q^2 = 0) &= 0.31 \pm 0.15\,, &
    A_0(q^2 = 0) &= 0.351 \pm 0.074\,, \nonumber \\
    A_1(q^2 = 0) &= 0.303 \pm 0.051\,, &
    A_{12}(q^2 = 0) &= 0.251 \pm 0.053\,, \nonumber \\
    T_1(q^2 = 0) &= 0.291 \pm 0.044\,, &
    T_2(q^2 = 0) &= 0.291 \pm 0.044\,, \nonumber \\
    T_3(q^2 = 0) &= 0.50 \pm 0.10\,. &
\end{align}

The total error can reach up to $\mathcal{O}(50 \%)$ at $q^2 = 0$ GeV$^2$ due to the extrapolation, but~it remains below 10\% in the low-recoil region.
\subsection{LCSR}
LCSRs for hadronic transition form factors are derived from a vacuum-to-hadron correlation function, with~two intermediate quark currents. In~given kinematical conditions, this correlation function can be expanded through an operator product expansion near the light-cone (LCOPE), in~terms of the hadron Light-Cone Distribution Amplitudes (LCDAs). For~$B \to K^{(*)}$ transitions, the hadron can be either the light meson (here $K^{(*)}$)~\cite{Braun:1999dp, Ball:2004rg, Ball:2004ye, Duplancic:2008tk, Bharucha:2015bzk,Khodjamirian:2017fxg} or the $B$-meson~\cite{Khodjamirian:2006st, Khodjamirian:2010vf, Lu:2018cfc, Cui:2022zwm, Gubernari:2018wyi, Monceaux:2023byy, Carvunis:2024koh, Descotes-Genon:2019bud, Descotes-Genon:2023ukb}. The~other hadron is then interpolated between the two quark currents which allows for the extraction of the form factor of interest. The~light-meson LCDAs are generally known with a better accuracy, and~higher-order corrections have been included. The~use of $B$-meson LCDAs is more recent and carries larger uncertainties. There have been discrepant predictions for the $B$-meson LCDAs~\cite{Wang:2015vgv, Braun:2003wx, Khodjamirian:2020hob, Nishikawa:2011qk, Rahimi:2020zzo} with significant implications as demonstrated in~\cite{Carvunis:2024koh}, with~an increase of up to $50 \%$ in $A_2^{B \to K^*}$ at $q^2 = 0$ GeV$^2$. Nonetheless, they allow for a direct computation of a large set of form factors for different processes. Using $B$-meson LCDAs makes it possible to take into account in addition the effects of the finite width of the $K^*$-meson. 

LCSRs rely on the semi-global quark--hadron duality~\cite{Colangelo:2000dp}, which induces a systematic error that is difficult to quantify. In~\cite{Carvunis:2024koh}, a~regime of the LCSR method was introduced to mitigate this error; however, conclusive results for these form factors are still~awaited.

\subsubsection[B -> K]{$B \to K$}

\begin{table}[H]
    \centering
    \begin{tabular}{|c|c|c|}
    \hline
         Form Factor & Value at $q^2 = 0$ & Ref. \\
         \hline \hline
         \multirow{ 4 }{*}{$f_+$} & $0.331 \pm 0.041 \pm \delta_{f_+}$$^\dagger$ &~\cite{Ball:2004ye}$^*$\\ 
         & $0.395 \pm 0.033$ &~\cite{Khodjamirian:2017fxg}$^*$ \\
         & $0.27 \pm 0.08$ &~\cite{Gubernari:2018wyi}$^{**}$ \\
         & $0.325 \pm 0.085$ &~\cite{Cui:2022zwm}$^{**}$ \\
         \hline
         \multirow{ 4 }{*}{$f_T$} &  $0.358 \pm 0.037 \pm \delta_{f_T}$$^\dagger$ &~\cite{Ball:2004ye}$^*$\\ 
         & $0.351 \pm 0.027$ &~\cite{Khodjamirian:2017fxg}$^*$ \\
         & $0.25 \pm 0.07$ &~\cite{Gubernari:2018wyi}$^{**}$ \\
         & $0.351 \pm 0.097$ &~\cite{Cui:2022zwm}$^{**}$ \\
         \hline
    \end{tabular}
    \caption{ LCSR predictions for $B \to K$ form factors at $q^2=0$, where $f_0(q^2=0) = f_+(q^2=0).$\\
    $^\dagger$: $\delta_{f_{+,T}}$ accounts for the uncertainty in the first Gegenbauer moment.\\
    $^*$: using $K$-meson LCDAs. \\ $^{**}$: using $B$-meson LCDAs.}
    \label{tab:FF_B_K}
\end{table}

In~\cite{Ball:2004ye}, the~form factors are obtained directly for $q^2 \leq 14$ GeV$^2$ before extrapolating to the entire physical range. The~results in~\cite{Khodjamirian:2017fxg} are expected to supersede those of~\cite{Ball:2004ye}, due to updates in inputs and the inclusion of higher-order corrections. In~\cite{Khodjamirian:2017fxg}, results are directly derived for $q^2 < 12$ GeV$^2$ and~extrapolated to the entire physical range using the BCL parameterisation~\cite{Bourrely:2008za}. The~LCSR results were later fitted in~\cite{Gubernari:2022hxn} by adding lattice points and using the parameterisation given in~\cite{Bharucha:2015bzk}.

The results in~\cite{Gubernari:2018wyi} are computed for $q^2 \leq 5$ GeV$^2$ and are then fitted with the $z$-expansion~\cite{Bharucha:2015bzk}, including additional points from the lattice determination~\cite{Bouchard:2013eph} to constrain the high-$q^2$ region. The~inclusion of lattice data does change the central values presented in Table~\ref{tab:FF_B_K}. In~a subsequent work~\cite{Gubernari:2022hxn}, the authors advise against using their previous results~\cite{Gubernari:2018wyi} for $B \to K$ due to issues with the determination of the sum rule thresholds that are not yet understood. The~authors of~\cite{Cui:2022zwm} work in the Soft-Collinear Effective Theory (SCET) framework, in~which they computed the next-order QCD corrections that have not been included in~\cite{Gubernari:2018wyi}. However, they do not necessarily supersede the results of~\cite{Gubernari:2018wyi} where the framework is the Heavy Quark Effective Theory (HQET) and~not SCET. Their results are obtained directly for $q^2 \leq q_{\text{cut}}^2$ where $q_{\text{cut}}^2$ varies between 8 GeV$^2$ and 10 GeV$^2$. The~extrapolation is performed using the BCL parameterisation~\cite{Bourrely:2008za}, including the lattice
results of~\cite{Parrott:2022rgu,Bailey:2015dka}.

\subsubsection[B -> K*]{$B \to K^*$}

\begin{table}[ht]
    \centering
    \begin{tabular}{|c|c|c|}
    \hline
         Form Factor & Value at $q^2 = 0$ & Ref. \\
         \hline \hline
         \multirow{ 4 }{*}{$A_1$} & $0.282 \pm 0.028 \pm \delta_{A_1}$$^\dagger$ &~\cite{Ball:2004rg}$^*$\\ 
         & $0.27 \pm 0.03$ &~\cite{Bharucha:2015bzk}$^*$ \\
         & $0.25^{+0.16}_{-0.10} $ &~\cite{Khodjamirian:2010vf}$^{**}$ \\
         & $0.26 \pm 0.08$ &~\cite{Gubernari:2018wyi}$^{**}$ \\
         \hline
         \multirow{ 3 }{*}{$A_2$} & $0.259 \pm 0.027 \pm \delta_{A_2}$$^\dagger$ &~\cite{Ball:2004rg}$^*$\\ 
         & $0.23^{+0.19}_{-0.10}$ &~\cite{Khodjamirian:2010vf}$^{**}$ \\
         & $0.24 \pm 0.09$ &~\cite{Gubernari:2018wyi}$^{**}$ \\
         \hline
         \multirow{ 1 }{*}{$A_{12}$}
         & $0.26 \pm 0.03$ &~\cite{Bharucha:2015bzk}$^*$ \\
         \hline
         \multirow{ 4 }{*}{$V$} & $0.411 \pm 0.033 \pm \delta_{V}$$^\dagger$ &~\cite{Ball:2004rg}$^*$\\ 
         & $0.34 \pm 0.04$ &~\cite{Bharucha:2015bzk}$^*$ \\
         & $0.36^{+0.23}_{-0.12}$ &~\cite{Khodjamirian:2010vf}$^{**}$ \\
         & $0.33 \pm 0.11$ &~\cite{Gubernari:2018wyi}$^{**}$ \\
         \hline
         \multirow{ 4 }{*}{$T_1$} & $0.333 \pm 0.028 \pm \delta_{T_1}$$^\dagger$ &~\cite{Ball:2004rg}$^*$\\ 
         & $0.28 \pm 0.03$ &~\cite{Bharucha:2015bzk}$^*$ \\
         & $0.31^{+0.18}_{-0.10}$ &~\cite{Khodjamirian:2010vf}$^{**}$ \\
         & $0.29 \pm 0.10$ &~\cite{Gubernari:2018wyi}$^{**}$ \\
         \hline
         \multirow{ 2 }{*}{$T_3$} & $0.202 \pm 0.018 \pm \delta_{T_3}$$^\dagger$ &~\cite{Ball:2004rg}$^*$\\ 
         & $0.22^{+0.17}_{-0.10}$ &~\cite{Khodjamirian:2010vf}$^{**}$ \\
         \hline
         \multirow{ 2 }{*}{$T_{23}$}
         & $0.67 \pm 0.08$ &~\cite{Bharucha:2015bzk}$^*$ \\
         & $0.58 \pm 0.13$ &~\cite{Gubernari:2018wyi}$^{**}$ \\
         \hline
    \end{tabular}
    \caption{LCSR predictions for $B \to K^*$ form factors at $q^2=0$ where $T_2(q^2=0) = T_1(q^2=0)$. The linear combinations $A_{12}$ and $T_{23}$ are introduced in Appendix~\ref{appendix:FFdef}.\\
    $^\dagger$: $\delta_{X}$ accounts for the uncertainty in the first Gegenbauer moment. \\
    $^*$: using $K^*$-meson LCDAs. \\ $^{**}$: using $B$-meson LCDAs.}
    \label{tab:FF_B_Kstar}
\end{table}

In~\cite{Ball:2004rg}, the~form factors are obtained directly for $q^2 \leq 14$ GeV$^2$ before being extrapolated to the whole physical range. These results are superseded in~\cite{Bharucha:2015bzk} due to the updated inputs. They are also determined directly for $q^2$ below $14$ GeV$^2$ and~fitted using their modified $z$-expansion (BSZ). The~numerical values presented in Table~\ref{tab:FF_B_Kstar} are the result of the fit using only their LCSR results. Additional results for a fit adding lattice points from~\cite{Horgan:2015vla} are also presented, which slightly shift the values presented in Table~\ref{tab:FF_B_Kstar}. 

In~\cite{Khodjamirian:2010vf}, the~form factors are obtained at $q^2 < 12$ GeV$^2$ and are then fitted using the BCL-parameterisation~\cite{Bourrely:2008za}. These results are superseded in~\cite{Gubernari:2018wyi} due to the updated inputs and the inclusion of higher order in the LCOPE. They are computed for $q^2 \leq 5$ GeV$^2$ and are then fitted with the BSZ $z$-expansion~\cite{Bharucha:2015bzk}, including additional points from the lattice determination~\cite{Horgan:2013hoa, Horgan:2015vla} to constrain the high-$q^2$ region. The~inclusion of lattice data does change the central values presented in Table~\ref{tab:FF_B_Kstar}. 

An advantage of LCSRs with $B$-meson LCDAs is the possibility of computing form factors for decays such as $B \to K \pi$, where the final state is a dimeson state, which allows us to study of the impact of the finite width of the $K^*$-meson. This computation has been carried out in~\cite{Descotes-Genon:2019bud} for the $P$-wave $K \pi$ system, and~then updated in~\cite{Descotes-Genon:2023ukb} to add the contribution of the $S$-wave $K \pi$. In~\cite{Descotes-Genon:2019bud}, a correction to the $B \to K^*$ form factors (using $B$-meson LCDAs) was obtained as a multiplicative factor of $\mathcal{W}_{K^*} \sim 1.1$. It corresponds to an enhancement in the decay rate $B \to K^* (\to K \pi) \ell \ell$ of $\mathcal{O}(20 \%)$ but has a negligible impact on $P_5'(B \to K^* (\to K \pi) \ell \ell)$. In~\cite{Descotes-Genon:2023ukb}, the additional $S$-wave corrections are claimed to be small. This $\mathcal{W}_{K^*} \sim 1.1$ correction is valid only in the low-$q^2$ region, where it has been computed, and~when using LCSR with $B$-meson~LCDAs. 

Since form factors are real analytic functions in the $q^2$ complex plane, up~to a branch cut and a pole at the resonance, it is customary to use parameterisations respecting these constraints, such as the Caprini, Lellouch, and~Neubert (CLN) parameterisation~\cite{Caprini:1997mu}; the~Boyd, Grinstein, and~Lebed (BGL) parameterisation~\cite{Boyd:1994tt}; the~Bourrely, Caprini, and~Lellouch (BCL) parameterisation~\cite{Bourrely:2008za}; the~Bharucha, Straub, and Zwicky (BSZ) parametrisation~\cite{Bharucha:2015bzk}; and the Gubernari, van Dyk, and Virto (GvDV) parameterisation~\cite{Gubernari:2020eft}. These parameterisations are used to extrapolate the results obtained for a limited range to the entire physical $q^2$ range. The~$z$-expansions are in practice truncated and only the first terms are considered for the fits, which induces a systematic truncation error. Dispersive (or unitarity) bounds allow for the control of said truncation errors for a given parameterisation and have been used for $B$-meson decays~\cite{deRafael:1992tu, deRafael:1993ib, Boyd:1994tt, Boyd:1995cf, Gubernari:2020eft, Gubernari:2022hxn, Gubernari:2023puw}. 

The latest dispersive bound results for local $B \to K^{(*)}$ form factors have been obtained in~\cite{Gubernari:2023puw} to which we refer for more~details.

\section{Non-Local~Contributions} \label{sec:non_local}
The long-distance effects in the $B \to K^{(*)} \ell \ell$ decays generated by four-quark and chromomagnetic dipole operators, sometimes referred to as charm-loop effects, are technically more challenging to derive than the local contributions. 
These long-distance effects were traditionally accounted for in the QCD factorisation framework and in the heavy quark limit for $q^2 < 7$~GeV$^2$. This calculation included up to weak annihilation, some non-factorisable contributions, and the hard spectator scattering~\cite{Beneke:2001at, Beneke:2004dp}. However, in~\cite{Gubernari:2022hxn}, it is suggested that this approach is reliable only below $4$ GeV$^2$ and may overlook potentially significant power corrections. Even outside of the resonance region, intermediate and/or virtual $c\bar{c}$ states still contribute. 
{An agnostic approach to take into account the effect of the power corrections is to consider a $q^2$- and transversity-dependent polynomial whose relative size is guesstimated~\cite{Hurth:2013ssa}.}

Two approaches have emerged in recent years to take into account these effects in a more comprehensive manner, which we refer to in the following as the $z$-expansion~\cite{Gubernari:2022hxn} and the hadronic dispersion relation~\cite{Lyon:2014hpa, Brass:2016efg, Blake:2017fyh, Cornella:2020aoq, Bordone:2024hui}. Both approaches start from a dispersion relation; however, they differ in their evaluation methods. The~$z$-expansion approach evaluates the dispersion relation using an LCOPE at negative $q^2$ before extrapolating it to the physical range. In~contrast, the~hadronic dispersion relation is evaluated directly at the hadronic level within the physical~range.

\subsection[The z-expansion]{The $z$-Expansion}
In the sum over accessible quark flavours in Equation~\eqref{eq:non_local_sum}, $\mathcal{H}_{u, \mu}^{B \to K^{(*)}}$ and $\mathcal{H}_{d, \mu}^{B \to K^{(*)}}$ are usually neglected as they are suppressed by CKM subleading matrix elements and/or small Wilson coefficients. The~necessary formulas for $\mathcal{H}_{s, \mu}^{B \to K^{(*)}}$ and $\mathcal{H}_{b, \mu}^{B \to K^{(*)}}$ have been derived in~\cite{Beneke:2001at, Beneke:2004dp, Asatrian:2019kbk} in the QCD factorisation framework and in the heavy quark mass limit, and~are given in Appendix C of~\cite{Gubernari:2022hxn}.

The most significant and challenging contributions of $\mathcal{H}_{c, \mu}^{B \to K^{(*)}}$, however, have been treated differently. 
In~the $z$-expansion approach~\cite{Gubernari:2022hxn}, the non-local contributions $\mathcal{H}_{c, \mu}^{B \to K^{(*)}}$ are computed at negative values of $q^2$ using an LCOPE. At~$q^2 = m_{J/\psi}^2$ these contributions are obtained from data on branching ratios and angular observables. This non-local form factor is then fitted via a $z$-expansion over the region $0 < q^2 < m_{J/\psi}^2$. 

The LCOPE computation was performed in~\cite{Gubernari:2020eft} considering only the dominant contributions from $\mathcal{O}_1^c$ and $\mathcal{O}_2^c$. In~\cite{Gubernari:2022hxn}, contributions from penguin operators were added, but~not from $\mathcal{O}_8$ as its contribution is considered negligible in this stage. The~LCOPE expansion~reads as follows:

\begin{equation}
\begin{aligned}
    \label{eq:non_local_LCOPE}
    \mathcal{H}_{c,\lambda}^{B \to K^{(*)}}
    &=
    -\frac{1}{16\pi^2}
    \left(
        \frac{q^2}{2m_B^2}
        \Delta C_9 \mathcal{F}^{B \to K^{(*)}}_{\lambda}
        + \frac{m_b}{m_B} \Delta C_7
        \mathcal{F}^{B \to K^{(*)}}_{T,\lambda}
    \right)
    +2 Q_c 
    \left(
        C_2 - \frac{C_1}{2 N_c}
    \right)
    \tilde{\mathcal{V}}^{B \to K^{(*)}}_{\lambda}
    \\&
    + \text{ higher-power corrections}
    \,,
\end{aligned}
\end{equation}
where the basis of local form factors $\mathcal{F}_{(T), \lambda}^{B \to K^{(*)}}$ is introduced in Appendix~\ref{appendix:alternate_basis}, the~matching coefficients $\Delta C_{7, 9}$ correspond to the leading power of non-local contributions, and $\tilde{\mathcal{V}}^{B \to K^{(*)}}_{\lambda}$ denotes subleading contributions that are not proportional to the local form factors. The~matching coefficients $\Delta C_{7, 9}$ have been computed to NLO in QCD~\cite{Asatryan:2001zw,Greub:2008cy,Ghinculov:2003qd,Bell:2014zya,deBoer:2017way,Asatrian:2019kbk} and~are sometimes incorporated into the effective Wilson coefficients $C_{7, 9}^{\text{eff}}$ which then become $q^2$-dependent. The~term $\tilde{\mathcal{V}}^{B \to K^{(*)}}_{\lambda}$ was initially found to be sizeable in~\cite{Khodjamirian:2010vf}, but~this computation has been superseded by that in~\cite{Gubernari:2020eft} where it was determined to be negligible due to a more complete calculation which included the missing three-particle distribution amplitudes and updated the necessary inputs. Thus, in~\cite{Gubernari:2022hxn}, to account for $\tilde{\mathcal{V}}^{B \to K^{(*)}}_{\lambda}$, only the uncertainty of the non-local form factors was~increased.

The LCOPE results and the residue extracted from the data at $q^2 = m_{J/\psi}^2$ are then fitted with the $z$ parameterisation described in~\cite{Gubernari:2020eft}. The~result is data-driven but~can also be used with only the LCOPE results for extrapolation, although this leads to larger~uncertainties.

Dispersive bounds for the non-local contributions were first derived in~\cite{Gubernari:2022hxn}. Imposing this dispersive bound for the long-distance effects significantly reduces the uncertainties. This dispersive bound can be further saturated by considering additional channels such as $\Lambda_b \to \Lambda \mu^+ \mu^-$. The~final results agree with the QCD factorisation approach; however, they exhibit larger uncertainties for the $z$-expansion method, especially near the $J/\psi$ pole, which is not accounted for in QCD factorisation. Assuming that the $z$-expansion accounts for the whole amplitude, a~careful assessment of the uncertainties is required to include all~contributions.

\subsection{The Hadronic Dispersion~Relation}

The evaluation of the long-distance effect in~\cite{Cornella:2020aoq} for $B \to K \ell \ell$ and its subsequent extension in~\cite{Bordone:2024hui} to $B \to K^{(*)} \ell \ell$ follows a slightly different and predominantly data-driven approach. This strategy can be viewed as an extension of the procedures suggested in~\cite{Lyon:2014hpa, Brass:2016efg, Blake:2017fyh}.
For~${B \to K \mu \mu}$~\cite{Cornella:2020aoq}, the long-distance effects are incorporated into the \mbox{effective Wilson coefficient:}
\begin{equation}
    C_9^{\text{eff}} = C_9 + Y_{c\bar{c}}(q^2)+ Y_{\text{light}}(q^2) + Y_{\tau \bar{\tau}}(q^2)\, ,
\end{equation}
where $Y_i(q^2)$ stands for the long-distance effect caused by the intermediate state $i$. Specifically, $Y_{c \bar{c}}$ accounts for the dominant contributions from $\mathcal{O}_1^c$ and $\mathcal{O}_2^c$ only, $Y_{\text{light}}$ corresponds to the contributions from the subleading operators $\mathcal{O}_{3-6, 8}$ and of $\mathcal{O}_{1-2}^u$, and~$Y_{\tau \bar{\tau}}$ corresponds to the tau loop. Instead of using an LCOPE to evaluate the dispersion relations at negative $q^2$, as~performed in~\cite{Khodjamirian:2010vf, Gubernari:2020eft, Gubernari:2022hxn}, the~dispersion relations here are evaluated directly at the hadronic level for positive $q^2$. For~$Y_{c \bar{c}}$, single-particle intermediate states ($J/\psi$, $\psi(2S)$, etc.) and two-particle states ($D\bar{D}$, $D^*\bar{D}$, etc.) are considered. For~$Y_{\text{light}}$, given the loop or CKM suppression and the inclusion of intermediate states with charm valence quarks in $Y_{c\bar{c}}$, only vector single-particle states with light valence quarks ($\rho$, $\omega$, $\phi$) are taken into account. The~hadronic contributions for single-particle intermediate states are modelled using Breit--Wigner distributions, while a more complex approximation is used for two-particle intermediate states. The~relevant amplitude and phase parameters are extracted from experimental measurements. The~dispersion relation for $Y_{c \bar{c}}$ is subtracted at $q^2 = 0$ to ensure convergence of the two-particle term, with~the term at $q^2 = 0$ evaluated in QCD factorisation. The~tau loop contribution is fully computed in perturbation~theory. 

For $B \to K^{*} \ell \ell$~\cite{Bordone:2024hui}, the~correction to $C_7$ is treated as a universal shift estimated using perturbation theory~\cite{Beneke:2001at}. The~corrections $Y(q^2)$ to $C_9$ were evaluated as
\begin{equation}
    Y^\lambda(q^2) = Y_{c\bar{c}}^\lambda(q^2) + Y_{q\bar{q}}^{\left[ 0\right]}(q^2) + Y_{b\bar{b}}^{\left[ 0\right]}(q^2)\,,
\end{equation}
where $\lambda$ denotes the polarisation. The~charm-loop contribution $Y^\lambda_{c\bar{c}}$ was evaluated hadronically, similar to the approach in~\cite{Cornella:2020aoq}, while the other contributions were taken from perturbation theory at the lowest order in $\alpha_s$. For~explicit expressions, we refer the reader to~\cite{Cornella:2020aoq}. 

The hadronic dispersion relation method has been implemented by LHCb in~\cite{LHCb:2016due, LHCb:2023gel, LHCb:2024onj}, revealing a persistent tension between Standard Model predictions and measurements. Overall, there is a strong agreement between the $z$-expansion and hadronic dispersion relation approaches. For~visual comparisons, we refer the reader to the plots in~\cite{LHCb:2024onj}.

In~\cite{Ciuchini:2022wbq}, the~impact of the rescattering of intermediate hadronic states is discussed. These contributions are particularly challenging to estimate. Based on a data-driven analysis, it is argued in~\cite{Ciuchini:2022wbq} that these contributions can potentially resolve the tension between SM predictions and experimental measurements, although~a consensus has not yet been reached. In~\cite{Isidori:2024lng}, an~estimate of the rescattering of charmed and charmed-strange mesons has been performed for the $B^0 \to K^0 \ell \ell$ decay, which finds at most a 10\% shift to $C_9$, insufficient to resolve the~tension.

\section{Other Sources of~Uncertainties} 
\label{sec:accuracy}
In this section, we discuss the impact of QED corrections on the Wilson coefficients and the influence of CKM matrix elements on predictions and their discrepancies with experimental measurements. These effects are often overlooked but are becoming increasingly relevant given the current precision of both measurements and~predictions.

\subsection{Wilson~Coefficients}
The Wilson coefficients are calculated in perturbative theory and  have been computed to NNLO in QCD~\cite{Bobeth:1999mk, Misiak:2004ew, Gorbahn:2004my}:
\begin{equation}
    C_i(\mu) = C_i^{(0)} + \frac{\alpha_s(\mu)}{4\pi}C_i^{(1)} + \Big( \frac{\alpha_s(\mu)}{4\pi} \Big)^2 C_i^{(2)} + \mathcal{O}(\alpha_s^3)\, ,
\end{equation}
where $C_i^{(n)}$ is the nth order of the Wilson coefficient in the $\alpha_s$ expansion. These coefficients are initially calculated at the electroweak scale $\mu_0 \sim M_W$ using a two-loop level computation, followed by the resummation of large logarithms and the evolution to the relevant scale $\mu_b \sim m_b$ using the three-loop Anomalous Dimension Matrix to account for operator mixing. Due to this operator mixing, it is customary to introduce the effective Wilson coefficients through the following combinations:
\begin{align}\label{eq:C7eff}
    C_7^{\text{eff}}(\mu) = C_7(\mu)  -  \frac{1}{3} C_3(\mu) - \frac{4}{9} C_4(\mu) - \frac{20}{3} C_5(\mu) - \frac{80}{9} C_6(\mu)\, , \\
    C_8^{\text{eff}}(\mu) = C_8(\mu)  + C_3(\mu) - \frac{1}{6} C_4(\mu) +20 C_5(\mu) - \frac{10}{3} C_6(\mu)\, .
\end{align}

We emphasise that the notation $C_7^{\text{eff}}$ used here arises solely from renormalisation and should not be confused with the use of effective Wilson coefficients in the context of non-local contributions, where ``effective'' refers to the inclusion of these non-local contributions within the Wilson~coefficients. 

QED corrections can also be computed and have been accounted for in~\cite{Bobeth:2003at, Huber:2005ig}. Introducing the variable $\kappa(\mu) = \frac{\alpha(\mu)}{\alpha_s(\mu)}$, the~perturbative expansion in both QED and QCD can be written as
\begin{equation}
    C_i(\mu) = \sum_{n, m = 0}^2 \alpha_s(\mu)^n \kappa(\mu)^m C_i^{(n, m)}(\mu) + \mathcal{O}(\alpha_s^3, \kappa^3)\, .
\end{equation}

Numerical values of the Wilson coefficients including NNLO QCD and NLO QED \footnote{We note that for $C_9$ and $C_{10}$ the QED corrections are said to be at NNLO in~\cite{Huber:2005ig}, but the definition of the operators $\mathcal{O}_9$ and $\mathcal{O}_{10}$ is different from the one used here.} corrections at the scale $\mu_b = 5$ GeV are reported in Table~\ref{tab:wilson_combined}.

\begin{table}[H]
    \centering
    \begin{tabular}{|c|c|c|}
    \hline
         Wilson Coefficient & Value (QCD) & Correction (QED) \\
         \hline \hline
        $C_1(\mu_b)$ & $-$0.2477 & $-$0.0030 \\ 
        $C_2(\mu_b)$ & 1.0080 & 0.0056 \\
        $C_3(\mu_b)$ & $-$0.0049 & $-$0.0000 \\
        $C_4(\mu_b)$ & $-$0.0763 & $-$0.0003 \\
        $C_5(\mu_b)$ & 0.0003 & 0.0000 \\
        $C_6(\mu_b)$ & 0.0009 & 0.0000 \\ 
        $C_7(\mu_b)$ & $-$0.3180 & 0.0037 \\
        $C_8(\mu_b)$ & $-$0.1710 & 0.0000 \\
        $C_9(\mu_b)$ & 4.1764 & $-$0.1305 \\
        $C_{10}(\mu_b)$ & $-$4.1494 & $-$0.1445 \\ \hline
    \end{tabular}
    \caption{Wilson coefficients at the scale $\mu_b = 5.0$ GeV. We use $\sin^2{\theta_W} = 0.231160$.}
    \label{tab:wilson_combined}
\end{table}

The QED contributions induce small corrections to the Wilson coefficients, up~to $\sim$3.5\% for $C_{9, 10}$ which are often neglected. Nevertheless, given the current precision, their inclusion is becoming increasingly~relevant. 
\subsection{CKM}
The determination of the CKM factor $\lambda_t = V_{tb}V_{ts}^*$ is particularly relevant for branching fractions, where the predictions scale with $\lvert V_{tb}V_{ts}^* \rvert ^2$. 

We use values from the Particle Data Group~\cite{ParticleDataGroup:2022pth} where the Wolfenstein parameter fit was performed with over-constraining measurements (and for which unitarity of the CKM matrix is implied):
\begin{align}
    &\lambda = 0.22500 \pm 0.00067\, , \hspace{0.5cm} A = 0.826^{+0.018}_{-0.015}\, , \nonumber \\
    & \bar{\rho} = 0.159 \pm 0.010\, , \hspace{1.5cm} \bar{\eta} = 0.348 \pm 0.010\, ,
\end{align}
which sets the following values:
\begin{equation}
    \lvert V_{ts} \rvert = 0.04110^{+0.00083}_{-0.00072}\, , 
    \lvert V_{tb} \rvert = 0.999118^{+0.000031}_{-0.000036}\, . 
\end{equation}
    
Very similar results were obtained in the \textit{CKM}fitter 2023 update~\cite{ValeSilva:2024jml} and the UTfit 2023 update~\cite{UTfit:2022hsi}, with~differences in the CKM factor $\lambda_t$ being well under a~percent. \\
While the uncertainties in the CKM matrix elements are relatively small compared to the uncertainty on local and non-local form factors, and~only amount to $\mathcal{O}(2 \%)$ for $\lvert V_{ts} \rvert$, they can still have a significant impact on the predictions due to the dependence on $\lvert V_{tb}V_{ts}^* \rvert ^2$. Between~2021 and 2022, the prediction of the $B \to K^{(*)} \mu \mu$ branching fraction has been shifted by $\mathcal{O}(5 \%)$ due to the update of the CKM values.
Moreover, if~we reconsider the assumptions of BSM physics or unitarity, the~CKM matrix elements are not immune to further shifts or changes. 
In addressing constraints in the presence of BSM physics, a~fit with tree-level inputs alone led to shifts of $\mathcal{O}(5 \%)$ and larger uncertainties in~\cite{ParticleDataGroup:2022pth}.

Therefore, given the precision of measurements and predictions, careful consideration of the CKM matrix elements and the assumptions under which they are obtained is crucial when discussing SM predictions. Notably, $P_5'$ serves as an ideal probe for NP due to its independence from CKM matrix elements by~construction.

\section{Impact on~Predictions}\label{sec:impact}
We discuss in this section the impact of local form-factors and long-distance effects on the branching fraction $\mathcal{B}(B^+ \to K^+ \mu \mu)$ and the angular observable $P_5'(B^0 \to K^{*0} \mu \mu)$ using the \texttt{SuperIso} public program~\cite{Mahmoudi:2007vz, Mahmoudi:2008tp, Neshatpour:2021nbn, Neshatpour:2022fak}. For~reference, we also plot the experimental results of LHCb~\cite{LHCb:2014cxe} and CMS~\cite{CMS:2024syx} for $\mathcal{B}(B^+ \to K^+ \mu \mu)$ and LHCb~\cite{LHCb:2020lmf} and CMS~\cite{CMS-PAS-BPH-21-002} for ${P_5'(B^0 \to K^{*0} \mu \mu)}$. The~grey vertical bands starting at $q^2 = 6$ GeV$^2$ denote the region approaching the resonances, which is less reliable and~sometimes disregarded. {The error bars on the figures in this section represent the $1\sigma$ errors computed by considering the propagated uncertainties across all parameters and inputs.}

\newpage
\subsection{Impact of Local Form Factors}

\begin{figure}[H]
    \hspace{-1.cm}
    \subfloat[]{\includegraphics[width=0.45\paperwidth]{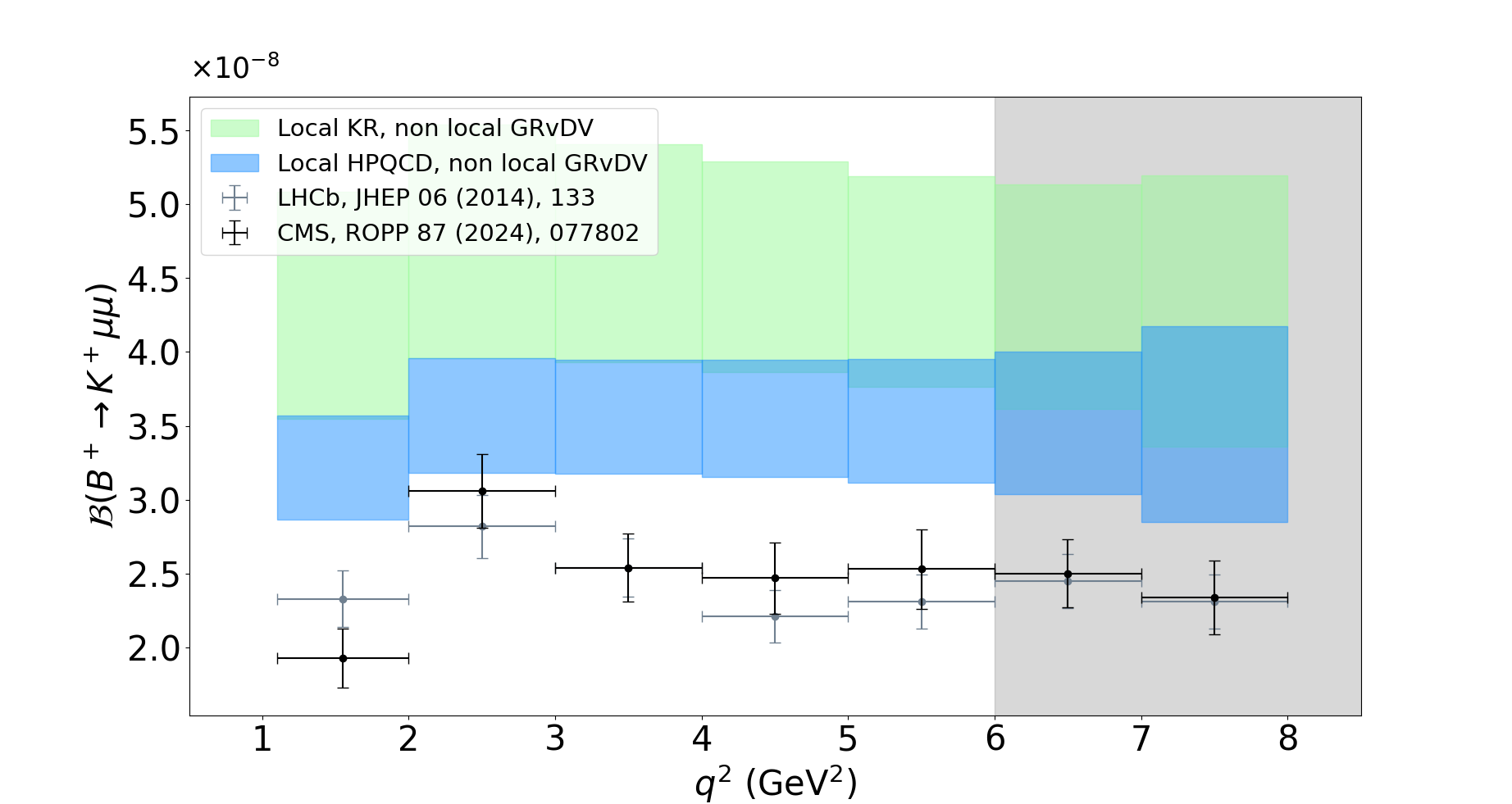}\label{fig:BR_local}}
    \subfloat[]{\includegraphics[width=0.45\paperwidth]{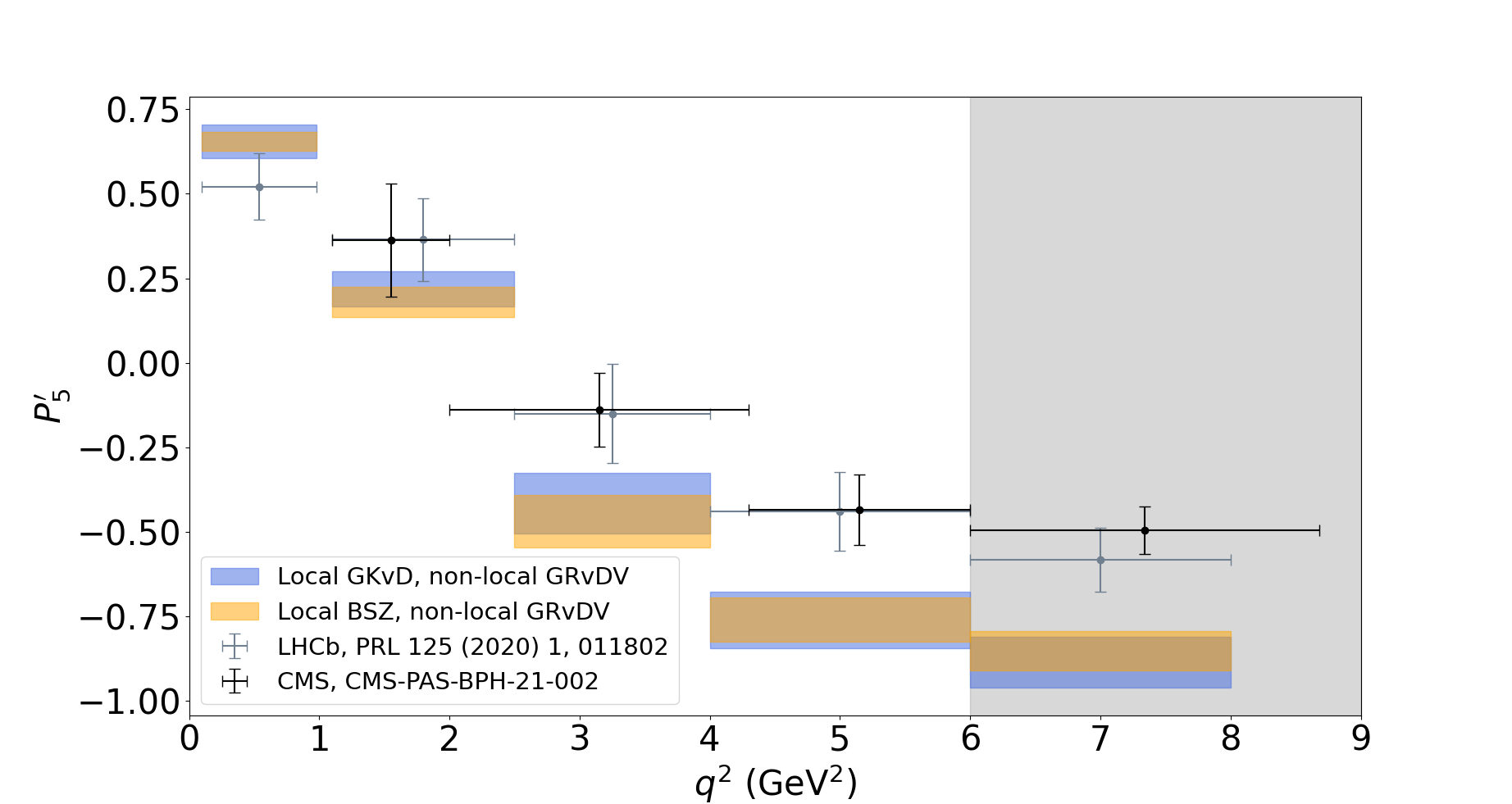}\label{fig:P'5_local}}
    \caption{Impact of local form factors on the prediction of (\textbf{a}) $\mathcal{B}(B^+ \to K^+ \mu \mu)$ with local form factors from~\cite{Khodjamirian:2017fxg} (KR) and from~\cite{Parrott:2022rgu} (HPQCD); (\textbf{b}) $P_5'(B_0 \to K^{*0} \mu \mu)$ with local form factors from~\cite{Gubernari:2018wyi} (GKvD) and from~\cite{Bharucha:2015bzk} (BSZ). For~both, non-local form factors from~\cite{Gubernari:2022hxn}, denoted as GRvDV, have been~used.}
    \label{fig:local}
\end{figure}

As expected, the~branching fraction $\mathcal{B}(B^+ \to K^+ \mu \mu)$ is significantly influenced by the local form factors. The~predictions shown in Figure \ref{fig:local}a, based on form factors from~\cite{Parrott:2022rgu}, agree within the error bars with those from~\cite{Khodjamirian:2017fxg}. However, the~observed tension with the measurements clearly depends on the specific set of form factors employed. Using the form factors from~\cite{Parrott:2022rgu}, for~instance, reveals a reduced tension. In~both cases, non-local contributions are adopted from~\cite{Gubernari:2022hxn}.

The optimised angular observables such as $P_5'$ are by construction less dependent on the local form factors, which is confirmed by the relatively small uncertainty on the predictions in Figure \ref{fig:local}b. Moreover the different determinations of the local form factors~\cite{Bharucha:2015bzk, Gubernari:2018wyi} are in agreement and consistent in their tension with the experimental data. 

\vspace{0.5cm}

\subsection{Impact of Non-Local Form Factors}

\begin{figure}[H]
    \hspace{-1.cm}
    \subfloat[\centering]{\includegraphics[width=0.45\paperwidth]{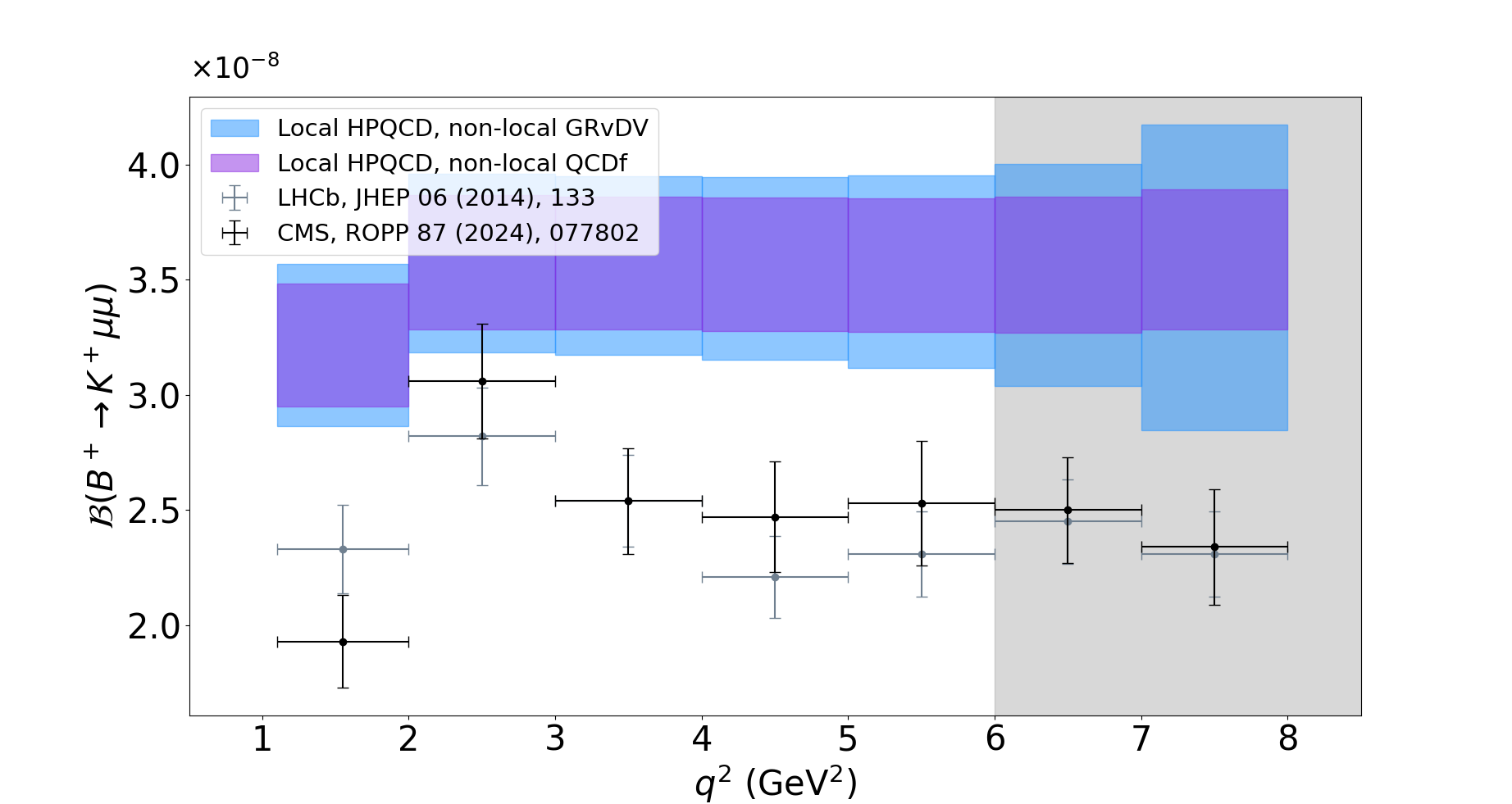}\label{fig:BR_non-local}}
    \subfloat[\centering]{\includegraphics[width=0.45\paperwidth]{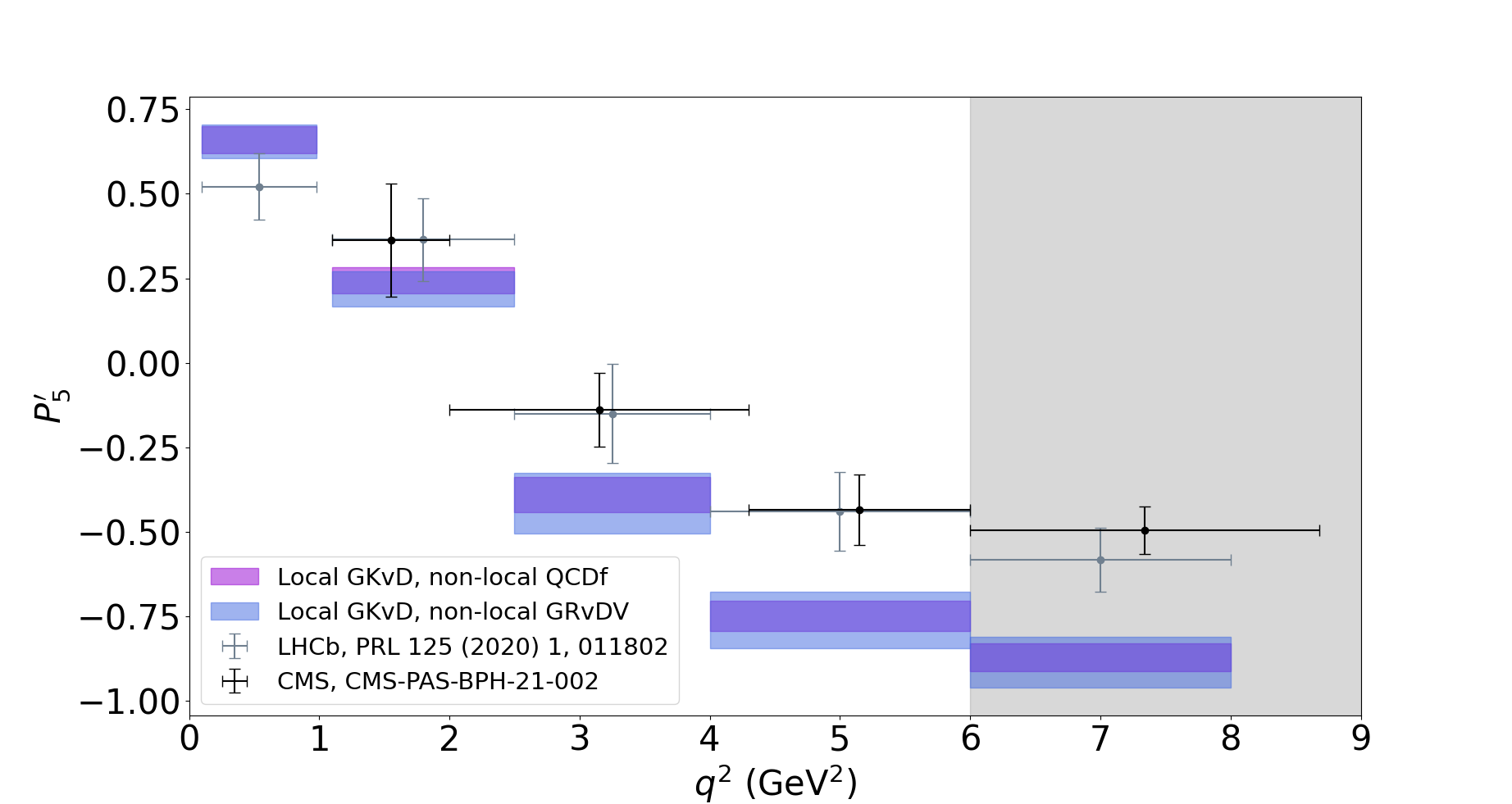}\label{fig:P'5_non-local}}
    \caption{Impact of non-local form factors on the prediction of (\textbf{a}) $\mathcal{B}(B^+ \to K^+ \mu \mu)$ with local form factors from~\cite{Parrott:2022rgu} (HQPCD), and~non-local form factors from~\cite{Beneke:2001at, Beneke:2004dp} (QCDf) and from~\cite{Gubernari:2022hxn} (GRvDV); (\textbf{b}) $P_5'(B_0 \to K^{*0} \mu \mu)$ with local form factors from~\cite{Gubernari:2018wyi} (GKvD), and~non-local form factors from~\cite{Beneke:2001at, Beneke:2004dp} (QCDf) and from~\cite{Gubernari:2022hxn} (GRvDV).}
    \label{fig:non-local}
\end{figure}

We compare the impact of non-local form factors obtained in QCD factorisation (QCDf)~\cite{Beneke:2001at, Beneke:2004dp} with those determined using the more recent $z$-expansion approach~\cite{Gubernari:2022hxn}. The~differences in central values for the branching fraction (Figure~\ref{fig:non-local}a) and the angular observable $P_5'$ (Figure~\ref{fig:non-local}b) are small in both implementations. However, since the QCDf approach does not account for charm resonances, there is a noticeable difference in uncertainties for the branching fraction. Consequently, in~the implementation of long-distance effects in \texttt{SuperIso}, an~error budget was guesstimated to account for the yet unknown power corrections when using QCDf results (see~\cite{Hurth:2013ssa,Hurth:2016fbr,Chobanova:2017ghn,Hurth:2017hxg,Arbey:2018ics,Hurth:2020rzx,Hurth:2023jwr} for more details).

\subsection{Impact of QED Corrections to Wilson Coefficients}

\begin{figure}[H]
    \hspace{-1.cm}
    \subfloat[\centering]{\includegraphics[width=0.45\paperwidth]{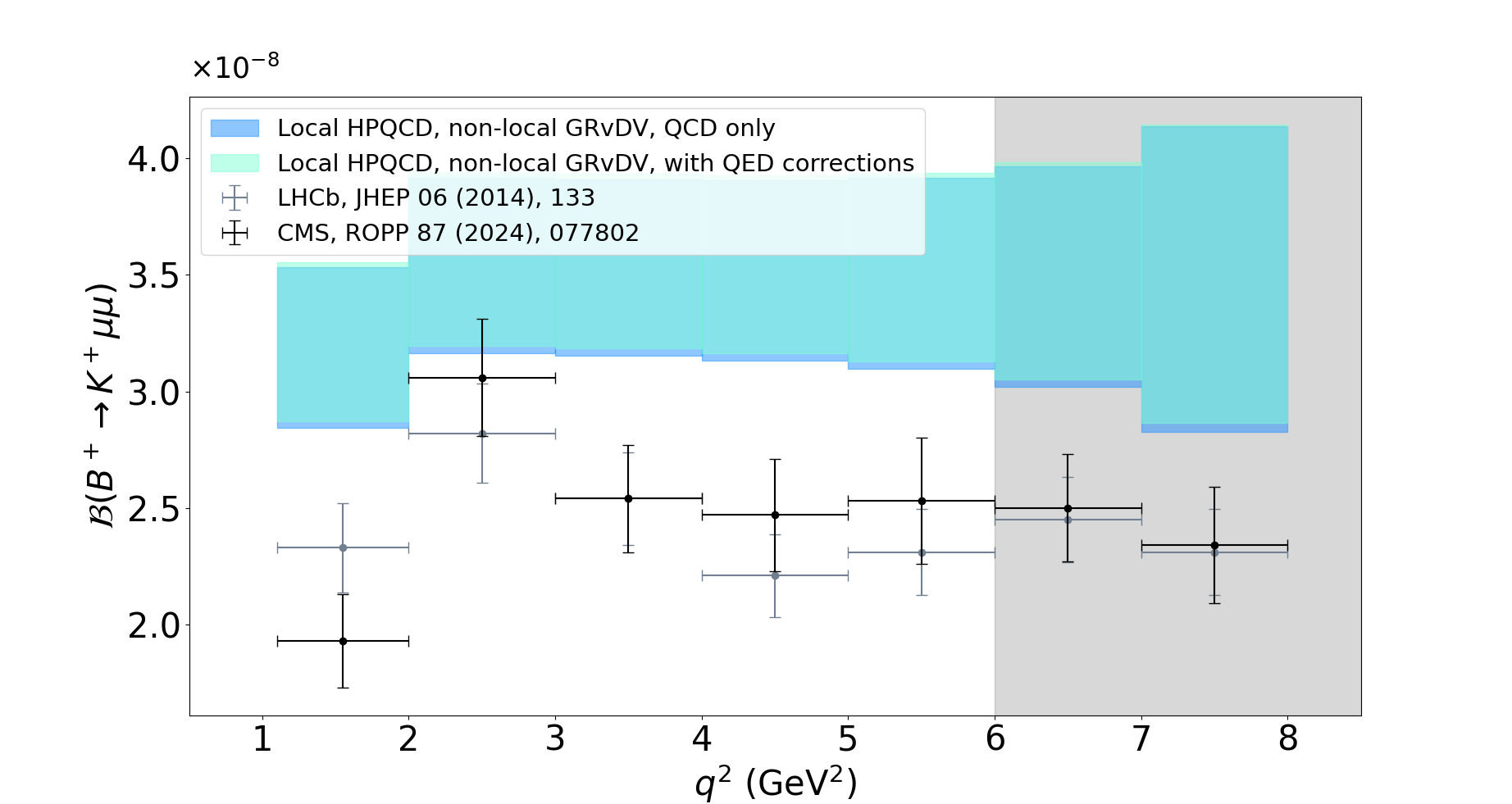}\label{fig:BR_QED}}
    \subfloat[\centering]{\includegraphics[width=0.45\paperwidth]{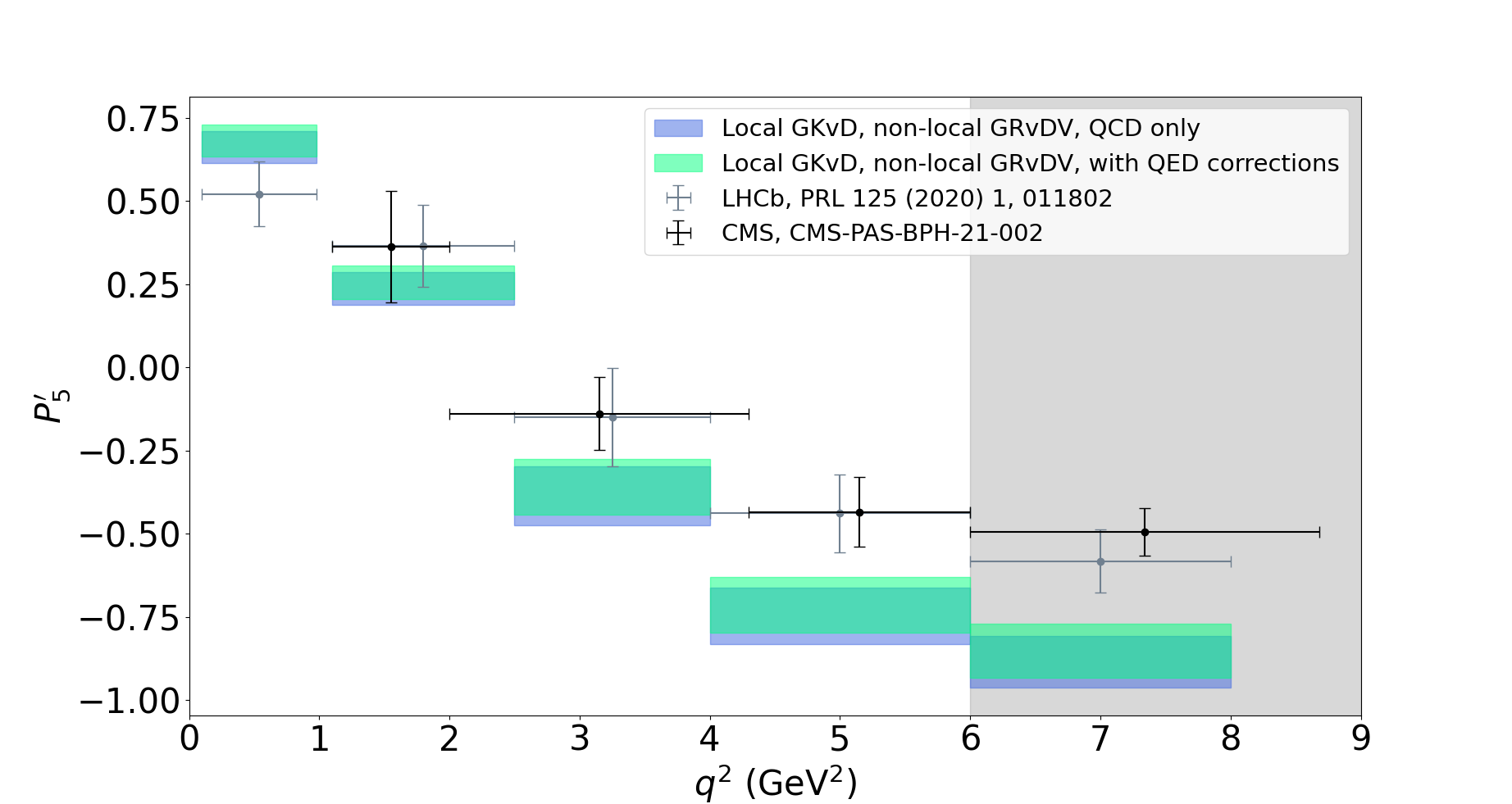}\label{fig:P'5_QED}}
    \caption{Impact of QED corrections to Wilson coefficients on the prediction of (\textbf{a}) $\mathcal{B}(B^+ \to K^+ \mu \mu)$ with local form factors from~\cite{Parrott:2022rgu} (HPQCD) and non-local form factors from~\cite{Gubernari:2022hxn} (GRvDV); (\textbf{b}) $P_5'(B_0 \to K^{*0} \mu \mu)$ with local form factors from~\cite{Gubernari:2018wyi} (GKvD) and non-local form factors from~\cite{Gubernari:2022hxn} (GRvDV).}
    \label{fig:QED}
\end{figure}

We compare the impact of QED corrections to Wilson coefficients. These corrections are relatively minor and the resulting shift observed in the branching fraction ${\mathcal{B}(B^+ \to K^+ \mu \mu)}$ in Figure \ref{fig:QED}a is negligible. However, their impact is more visible in $P_5'$ as shown in Figure~\ref{fig:QED}b, where they tend to reduce the tension between predictions and~measurements.

\subsection{Impact of CKM Matrix Elements}

\begin{figure}[H]
    \centering
\includegraphics[width=0.42\paperwidth]{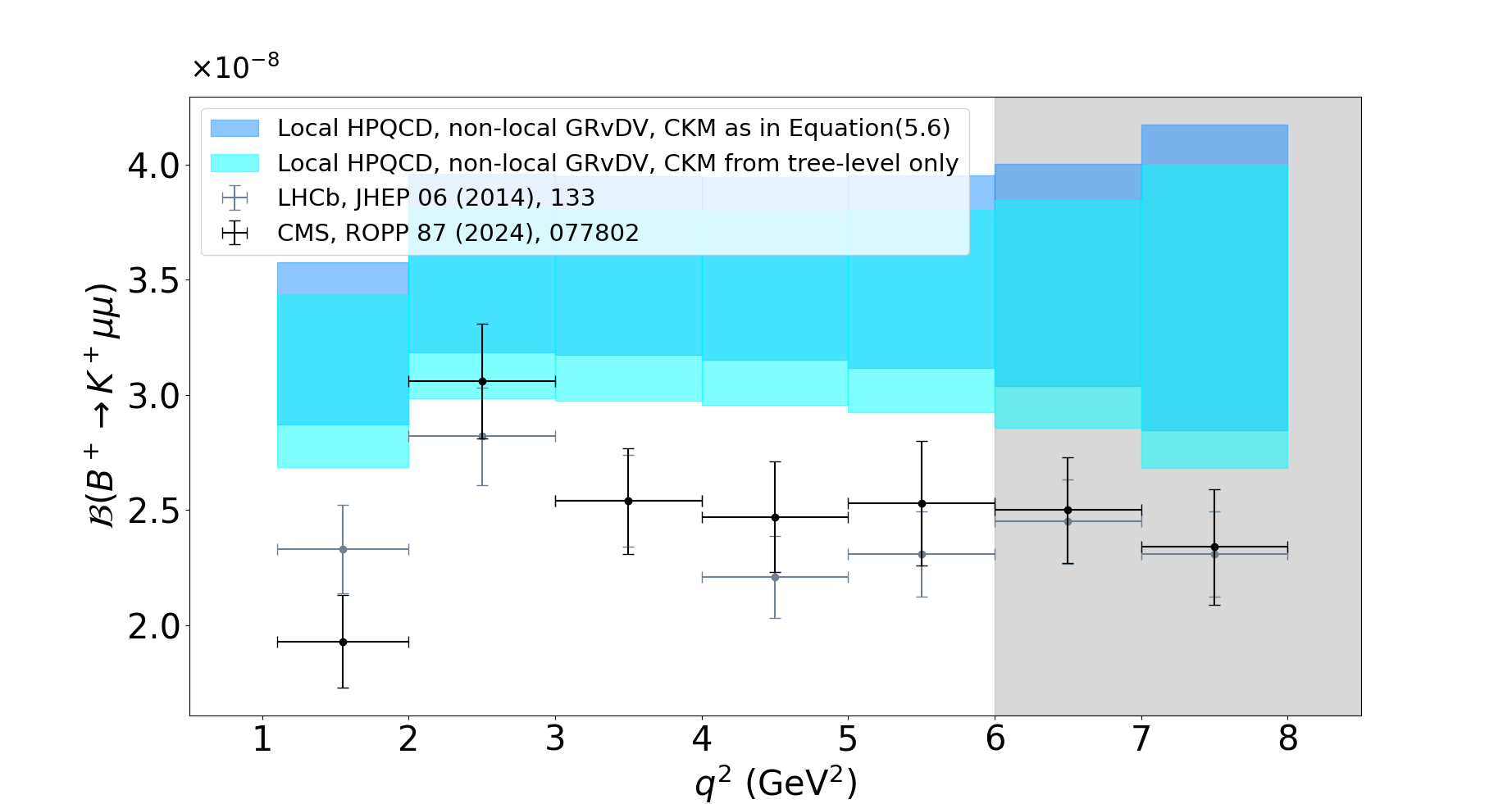}
    \hspace{-5pt}\caption{Impact
 of the CKM factor on the prediction of ${\mathcal{B}(B^+ \to K^+ \mu \mu)}$ with local form factors from~\cite{Parrott:2022rgu} (HPQCD) and non-local form factors from~\cite{Gubernari:2022hxn} (GRvDV).}
    \label{fig:BR_CKM}
\end{figure}

We compare below the impact of the CKM matrix elements obtained from the fits in~\cite{ParticleDataGroup:2022pth}: One with the full data, and~one taking only into account tree-level inputs (Figure~\ref{fig:BR_CKM}). There is a clear effect from the set of CKM parameters used when discussing the tension with the experimental~data.

\unskip

\section{Conclusions}\label{sec:Conclusion}
Hints of NP beyond the SM in semi-leptonic $B$ decays, particularly in the branching fraction $\mathcal{B}(B \to K \mu \mu)$ and the angular observable $P_5'(B \to K^* \mu \mu)$ at low $q^2$, still persist. However, unlike the $R_{K^{(*)}}$ ratios, accurately assessing the level of tension with experimental measurements for these observables is significantly more challenging. In~this review, we addressed the theoretical calculations of $B \to K^{(*)} \ell \ell$ and~the main sources of~uncertainties.

The primary elements contributing to the remaining uncertainty in $B \to K^{(*)} \ell \ell$ decays are the local transition form factors and the long-distance contributions. Significant progress has been achieved in lattice QCD, particularly by the HPQCD collaboration, which has provided direct results for $f_{+, T}^{B \to K}$ across the entire $q^2$ range. Nevertheless, as~we await similar results for $B \to K^*$, and~a cross-check for $B \to K$, LCSRs remain relevant. Despite all predictions for the relevant form factors being consistent within their uncertainties, they significantly influence the predictions, with~the tension reduced when using the HPQCD form factor set. Optimised angular observables such as $P_5'$ are less sensitive to local form factors, and~furthermore, the~discrepancies among various determinations of local $B \to K^*$ form factors are smaller. However, it is crucial to note that LCSRs with $B$-meson LCDAs heavily rely on $B$-LCDA parameters, which currently exhibit discrepancies and can lead to substantial variations in the local form factors at $q^2 = 0$ GeV$^2$, up~to $\mathcal{O}(50\%)$. 

Both types of observables are impacted by the non-local effects, although~different estimates of these effects are in agreement. While the QCD factorisation framework initially provided an estimate of these non-local effects, more recent methods have emerged such as the $z$-expansion, which computes charm-loop contributions using an LCOPE at negative $q^2$, and~the hadronic dispersion relation, evaluated predominantly at the hadronic level within the physical range, and~have yielded consistent results. 
{It is worth noting that, in~this stage, the~different evaluations of the non-local contributions do not resolve the discrepancies between theoretical predictions and experimental results; nonetheless, it is not excluded that some contributions (e.g., the rescattering of intermediate hadronic states) with potentially significant effects have been overlooked.}

Finally, we discussed the impact of QED corrections on the Wilson coefficients. While their effect on branching fractions is negligible, this is less the case for angular observables. Furthermore, careful consideration is warranted regarding the CKM matrix elements, given that the assumptions used to determine them have a substantial impact on branching fractions and other observables that are sensitive to~them.

In recent years, there have been remarkable improvements in the predictions of $B \to K^{(*)} \ell \ell$, and~we expect this progress to continue, leading to increasingly precise predictions, and~further constraints or hints on New Physics~signatures.

\vspace{6pt} 

\paragraph{Acknowledgement:} 
The authors are grateful to A. Carvunis, S. Neshatpour, and T. Hurth for useful~discussions.
This research was funded in part by the National Research Agency (ANR) under project ANR-21-CE31-0002-01.

\appendix
\section[\appendixname~\thesection]{Definition of Local Hadronic form Factors}  \label{appendix:FFdef}
\renewcommand{\theequation}{A\arabic{equation}} 
\setcounter{equation}{0} 

A suitable form factor basis for the $B \to K$ transition is $f^{B \to K}_0$, $f^{B \to K}_+$, and $f^{B \to K}_T$, which are defined as
\begin{equation}
\begin{aligned}
\left\langle K(k)\left|\bar{q}_1 \gamma^\mu b\right| B(p_B)\right\rangle & =\left[(p_B+k)^\mu-\frac{m_B^2-m_P^2}{q^2} q^\mu\right] f_{+}^{B \rightarrow K}+\frac{m_B^2-m_P^2}{q^2} q^\mu f_0^{B \rightarrow K}\,, \\
\left\langle K(k)\left|\bar{q}_1 \sigma^{\mu \nu} q_\nu b\right| B(p_B)\right\rangle & =\frac{i f_T^{B \rightarrow K}}{m_B+m_P}\left[q^2(p_B+k)^\mu-\left(m_B^2-m_P^2\right) q^\mu\right]\,.
\end{aligned}
\end{equation}

For $\bar{B} \to\bar{ K}^*$, we used the form factors $V^{B \rightarrow K^*}$, $A_0^{B \rightarrow K^*}$, $A_1^{B \rightarrow K^*}$, $A_2^{B \rightarrow K^*}$, $T_1^{B \rightarrow K^*}$, $T_2^{B \rightarrow K^*}$, and $T_3^{B \rightarrow K^*}$, which can be defined as

\begin{equation}
\begin{aligned}
\left\langle K^*(k, \eta)\left|\bar{q}_1 \gamma^\mu b\right| B(p_B)\right\rangle = & \epsilon^{\mu \nu \rho \sigma} \eta_\nu^* p_{B\rho} k_\sigma \frac{2 V^{B \rightarrow K^*}}{m_B+m_{K^*}}\,, \\
\left\langle K^*(k, \eta)\left|\bar{q}_1 \gamma^\mu \gamma_5 b\right| B(p_B)\right\rangle = & i \eta_\nu^*\left[q^\mu q^\nu \frac{2 m_{K^*}}{q^2}A_0^{B \rightarrow K^*} + \left(g^{\mu \nu} - \frac{q^\mu q^\nu}{q^2}\right)\left(m_B+m_{K^*}\right) A_1^{B \rightarrow K^*} \right.\\
& -\left(\frac{(p_B+k)^\mu q^\nu}{m_B+m_{K^*}} - \frac{q^\mu q^\nu}{q^2}(m_B - m_{K^*})\right)A_2^{B \rightarrow K^*}  \biggr]\,, \\
\left\langle K^*(k, \eta)\left|\bar{q}_1 i \sigma^{\mu \nu} q_\nu b\right| B(p_B)\right\rangle 
 = & - \epsilon^{\mu \nu \rho \sigma} \eta_\nu^* p_{B\rho} k_\sigma 2 T_1^{B \rightarrow K^*}\,, \\
\left\langle K^*(k, \eta)\left|\bar{q}_1 i \sigma^{\mu \nu} q_\nu \gamma_5 b\right| B(p_B)\right\rangle = & i \eta_\nu^*\left[\left(g^{\mu \nu}\left(m_B^2-m_{K^*}^2\right)-(p_B+k)^\mu q^\nu\right) T_2^{B \rightarrow K^*}\right. \\
 & + q^\nu\left(q^\mu-\frac{q^2}{m_B^2-m_{K^*}^2}(p_B+k)^\mu\right) T_3^{B \rightarrow K^*}\bigr]\,,
\end{aligned}
\end{equation}

The variables $p_B$, $k$, and~$q$ represent the momenta of the $B$-meson, the~$K^{(*)}$-meson, and~the momentum transfer, respectively. $\eta$ denotes the polarisation of the $K^*$-meson. We adopt the convention $\epsilon_{0123} = +1$.

It is convenient to introduce the following linear combination of form factors $A_{12}$ and $T_{23}$ as 

\begin{align}
    A_{12} &\equiv \frac{(m_B+m_{K^*})^2(m_B^2 - m_{K^*}^2 - q^2)A_1 - \lambda(m_B^2, m_{K^*}^2, q^2) A_2}{16m_Bm_{K^*}^2(m_B + m_{K^*})}\,, \\
    T_{23} &\equiv \frac{(m_B^2-m_{K^*}^2)(m_B^2 + 3 m_{K^*}^2 - q^2)T_2 - \lambda(m_B^2, m_{K^*}^2, q^2) T_3}{8m_Bm_{K^*}^2(m_B - m_{K^*})}\,.
\end{align}

\section[\appendixname~\thesection]{Angular Conventions} \label{appendix:angular}
\renewcommand{\theequation}{B\arabic{equation}} 
\setcounter{equation}{0} 

We collect below more precise definitions of the angles relevant to the $\bar{B} \to \bar{K}^* \ell \ell$ decay. Let us define the symmetric and antisymmetric momenta:
\begin{equation}
\begin{split}
    \begin{aligned}
        \overrightarrow{P}_{\ell^-\ell^+}^i &= \overrightarrow{p_{\ell^-}} + \overrightarrow{p_{\ell^+}}\,, \\
        \overrightarrow{P}_{K\pi}^i &= \overrightarrow{p_{K}} + \overrightarrow{p_{\pi}}\,,
    \end{aligned}
    \qquad\qquad
    \begin{aligned}
        \overrightarrow{Q}_{\ell^-\ell^+}^i &= \overrightarrow{p_{\ell^-}} - \overrightarrow{p_{\ell^+}}\,, \\
        \overrightarrow{Q}_{K\pi}^i &= \overrightarrow{p_{K}} - \overrightarrow{p_{\pi}}\,,
    \end{aligned}
\end{split}
\end{equation}
where the superscript $i$ indicates in which particle's rest frame the momenta are~evaluated. 

The angles defined with the convention in Section~\ref{sec:B_to_Kstar_l_l} can then be expressed as

\begin{equation}
\begin{split}
    \begin{aligned}
        \cos{\theta_\ell} &= \frac{\overrightarrow{Q}_{\ell^-\ell^+}^{\ell\ell}.\overrightarrow{P}_{K\pi}^{\ell\ell}}{|\overrightarrow{Q}_{\ell^-\ell^+}^{\ell\ell}||\overrightarrow{P}_{K\pi}^{\ell\ell}|}\,, \\ 
        \cos{\theta_K} &= - \frac{\overrightarrow{Q}_{K\pi}^{K^*}.\overrightarrow{P}_{\ell^-\ell^+}^{K^*}}{|\overrightarrow{Q}_{K\pi}^{K^*}||\overrightarrow{P}_{\ell^-\ell^+}^{K^*}|}\,, \\ 
    \end{aligned}
    \qquad
    \begin{aligned}
        \cos{\phi} &= \frac{(\overrightarrow{Q}_{\ell^-\ell^+}^{\bar{B}} \times \overrightarrow{P}_{\ell^-\ell^+}^{\bar{B}})}{|\overrightarrow{Q}_{\ell^-\ell^+}^{\bar{B}}|| \overrightarrow{P}_{\ell^-\ell^+}^{\bar{B}}|}.\frac{(\overrightarrow{Q}_{K\pi}^{\bar{B}} \times \overrightarrow{P}_{K\pi}^{\bar{B}})}{|\overrightarrow{Q}_{K\pi}^{\bar{B}}|| \overrightarrow{P}_{K\pi}^{\bar{B}}|}\,,  \\
        \sin{\phi} &= \bigg(\frac{(\overrightarrow{Q}_{\ell^-\ell^+}^{\bar{B}} \times \overrightarrow{P}_{\ell^-\ell^+}^{\bar{B}})}{|\overrightarrow{Q}_{\ell^-\ell^+}^{\bar{B}}|| \overrightarrow{P}_{\ell^-\ell^+}^{\bar{B}}|}\times\frac{(\overrightarrow{Q}_{K\pi}^{\bar{B}} \times \overrightarrow{P}_{K\pi}^{\bar{B}})}{|\overrightarrow{Q}_{K\pi}^{\bar{B}}|| \overrightarrow{P}_{K\pi}^{\bar{B}}|} \bigg). \frac{\overrightarrow{P}_{K\pi}^{\bar{B}}}{|\overrightarrow{P}_{K\pi}^{\bar{B}}|}\,. \\
    \end{aligned}
\end{split}
\end{equation}

We refer to~\cite{Gratrex:2015hna} for a detailed discussion on the different conventions, both theoretical and the ones used by LHCb, and~how to convert between~them.

\section[\appendixname~\thesection]{Alternate Basis}\label{appendix:alternate_basis}
We relate the usual form factor basis that we use to the one presented in~\cite{Gubernari:2022hxn}, as~well as the equivalence between the $F_i$ functions for $B \to K \ell \ell$ and the transversity amplitudes for ${\bar{B} \to \bar{K}^* \ell \ell}$ with the transversity amplitudes in~\cite{Gubernari:2022hxn}.

\subsection[\appendixname~\thesubsection]{$B \to K$}
\renewcommand{\theequation}{C1.\arabic{equation}} 
\setcounter{equation}{0} 

The transversity amplitudes introduced in~\cite{Gubernari:2022hxn} for the $B \to K \ell \ell$ transition read as follows:
\begin{align}\label{eq:BK_transversity}
    \mathcal{A}^{K\ell\ell}_{\lambda,L(R)}
        & = \mathcal{N}^{K\ell\ell} \left\lbrace(C_9 \mp C_{10}) \mathcal{F}_{\lambda}^{B \to K}
          + \frac{2 m_b m_B}{q^2} \left[C_7 \mathcal{F}_{T,\lambda}^{B \to K} - 16 \pi^2 \frac{m_B}{m_b} \mathcal{H}_{\lambda}^{B \to K} \right]\right\rbrace\,,
\end{align}
where $\mathcal{N}^{K\ell\ell} = \sqrt{\lambda(m_B^2, m_K^2, q^2) \times \mathcal{C}(q^2)}$ and $\lambda=0,\,t$ refers to the longitudinal or timelike polarisation, respectively, with~$\mathcal{F}_{T,t}^{B \to K} = \mathcal{H}_{t}^{B \to K} =0 $.
The form factor basis in~\cite{Gubernari:2022hxn} is related to the one we use and introduced in Appendix~\ref{appendix:FFdef} by
\begin{equation}
\begin{split}
    \begin{aligned}
        \mathcal{F}_{0}   &= f_+^{B\to P}\,, \\
        \mathcal{F}_{t}   &= f_0^{B\to P}\,,
    \end{aligned}
    \qquad\qquad
    \begin{aligned}
    &\mathcal{F}_{T,0} = \frac{q^2}{m_B(m_B+m_K)} f_T^{B\to P}\,.
    \end{aligned}
\end{split}
\end{equation}

The $F_i$ functions introduced in Section~\ref{sec:B_to_K_l_l} are related to the transversity amplitudes~\eqref{eq:BK_transversity} as:
\begin{align}
    F_V(q^2) &= \frac{1}{\mathcal{N}^{K\ell\ell}} \frac{\mathcal{A}_{0,L}+\mathcal{A}_{0,R}}{2}\,, \hspace{2.5cm} F_A(q^2) = \frac{1}{\mathcal{N}^{K\ell\ell}} \frac{\mathcal{A}_{0,R}-\mathcal{A}_{0,L}}{2}\,, \notag\\
         F_P(q^2) &= -\frac{m_\ell}{\mathcal{N}^{K\ell\ell}} \Big(\frac{\mathcal{A}_{0,R}-\mathcal{A}_{0,L}}{2} + \frac{m_B^2 - m_K^2}{q^2} \bigg(\frac{\mathcal{A}_{t,L}-\mathcal{A}_{t,R}}{2} +\frac{\mathcal{A}_{0,R}-\mathcal{A}_{0,L}}{2}\bigg)\Big)\,,
\end{align}
{and \begin{equation}
    \delta F_V = -\frac{32 \pi^2 m_B^2}{q^2} \mathcal{H}_0^{B \to K}\,.
\end{equation}}
\subsection[\appendixname~\thesubsection]{$B \to K^*$}
\renewcommand{\theequation}{C2.\arabic{equation}} 
\setcounter{equation}{0} 
The transversity amplitudes introduced in~\cite{Gubernari:2018wyi} for the $B \to K^* \ell \ell$ transition read as follows:
\begin{align}
    \mathcal{A}^{K^*\ell\ell}_{\lambda,L(R)}
        & = \mathcal{N}^{K^*\ell\ell} \left\lbrace(C_9 \mp C_{10}) \mathcal{F}^{B\to K^*}_{\lambda} + \frac{2 m_b m_B}{q^2} \left[C_7 \mathcal{F}^{B\to K^*}_{T,\lambda} - 16 \pi^2 \frac{m_B}{m_b} \mathcal{H}^{B\to K^*}_{\lambda} \right]\right\rbrace
        \,,
    \label{eq:tampBKstar}
\end{align}
where $\mathcal{N}^{K^* \ell \ell} = N \times m_B$ and $\lambda=\perp,\,\parallel, 0, t$ refers to the different polarisations, with~$\mathcal{F}^{B\to K^*}_{T,t} =\mathcal{H}^{B\to K^*}_{t} =0 $.
We further introduce $\mathcal{A}_t^{K^* \ell \ell} = \mathcal{A}_{t, L}^{K^* \ell \ell} - \mathcal{A}_{t,R}^{K^* \ell \ell}$.
Their form factor basis is related to the one we use and introduced in Appendix~\ref{appendix:FFdef} by

\begin{align}
    \mathcal{F}_{\perp}        & = \frac{\sqrt{2\,\lambda(m_B^2, m_{K^*}^2, q^2)}}{m_B (m_B + m_{K^*})} V\,, \hspace{1.cm} \mathcal{F}_{\parallel} = \frac{\sqrt{2} (m_B + m_{K^*})}{m_B} A_1\,,
    \notag \\
    \mathcal{F}_{0}            & = \frac{(m_B^2 - m_{K^*}^2 - q^2)(m_B + m_{K^*})^2 A_1 - \lambda(m_B^2, m_{K^*}^2, q^2) A_2}{2 m_{K^*} m_B^2 (m_B + m_{K^*}) }\,, \notag
    \\
    \mathcal{F}_{t}            & = A_0\,, \hspace{0.5cm}  \mathcal{F}_{T,\perp}   = \frac{\sqrt{2\, \lambda(m_B^2, m_{K^*}^2, q^2)}}{m_B^2} T_1\,, \hspace{0.5cm}   \mathcal{F}_{T,\parallel}   = \frac{\sqrt{2} (m_B^2 - m_{K^*}^2)}{m_B^2} T_2\,,
 \notag   \\
    \mathcal{F}_{T,0}      & = \frac{q^2 (m_B^2 + 3 m_{K^*}^2 - q^2)}{2 m_B^3 m_{K^*}} T_2
                                - \frac{q^2\lambda(m_B^2, m_{K^*}^2, q^2)}{2 m_B^3 m_{K^*} (m_B^2 - m_{K^*}^2)} T_3\,.
\end{align}

The transversity amplitudes $A^{K^*\ell\ell}_{\lambda,L(R)}$ introduced in Section~\ref{sec:B_to_Kstar_l_l} are related to the transversity amplitudes $\mathcal{A}^{K^*\ell\ell}_{\lambda,L(R)}$ by
\begin{equation}
\begin{aligned}
    A^{K^*\ell\ell}_{\perp,L(R)} & = \mathcal{A}^{K^*\ell\ell}_{\perp,L(R)}\,,
    &
    A^{K^*\ell\ell}_{\parallel,L(R)} & = - \mathcal{A}^{K^*\ell\ell}_{\parallel,L(R)}\,,
    &
    \\
    A^{K^*\ell\ell}_{0,L(R)} & = - \frac{m_B}{\sqrt{q^2}}\mathcal{A}^{K^*\ell\ell}_{0,L(R)}\,,
    &
    A^{K^*\ell\ell}_{t} & = - \frac{1}{m_B}\sqrt{\frac{\lambda(m_B^2, m_{K^*}^2, q^2)}{q^2}}\mathcal{A}^{K^*\ell\ell}_{t}\,.
    &
\end{aligned}
\end{equation}

Similarly,
\begin{equation}
\begin{aligned}
    \delta A^{K^*\ell\ell}_{\perp,L(R)} & = - 32 \pi^2 N \frac{m_B^3}{q^2}\mathcal{H}^{K^*\ell\ell}_{\perp,L(R)}\,,
    &
    \delta A^{K^*\ell\ell}_{\parallel,L(R)} & = + 32 \pi^2 N \frac{m_B^3}{q^2}\mathcal{H}^{K^*\ell\ell}_{\parallel,L(R)}\,,
    &
    \\
    \delta A^{K^*\ell\ell}_{0,L(R)} & = + 32 \pi^2 N \frac{m_B^4}{q^2\sqrt{q^2}}\mathcal{H}^{K^*\ell\ell}_{0,L(R)}\,.
    &
\end{aligned}
\end{equation}

\newpage
\addcontentsline{toc}{section}{References}
\bibliographystyle{JHEP}
\bibliography{bibli}

\providecommand{\href}[2]{#2}\begingroup\raggedright\begin{thebibliography}{100}

\bibitem{CMS:2012qbp}
{\scshape CMS} collaboration, \emph{{Observation of a New Boson at a Mass of 125 GeV with the CMS Experiment at the LHC}}, \href{https://doi.org/10.1016/j.physletb.2012.08.021}{\emph{Phys. Lett. B} {\bfseries 716} (2012) 30} [\href{https://arxiv.org/abs/1207.7235}{{\ttfamily 1207.7235}}].

\bibitem{ATLAS:2012yve}
{\scshape ATLAS} collaboration, \emph{{Observation of a new particle in the search for the Standard Model Higgs boson with the ATLAS detector at the LHC}}, \href{https://doi.org/10.1016/j.physletb.2012.08.020}{\emph{Phys. Lett. B} {\bfseries 716} (2012) 1} [\href{https://arxiv.org/abs/1207.7214}{{\ttfamily 1207.7214}}].

\bibitem{LHCb:2022vje}
{\scshape LHCb} collaboration, \emph{{Measurement of lepton universality parameters in $B^+\to K^+\ell^+\ell^-$ and $B^0\to K^{*0}\ell^+\ell^-$ decays}}, \href{https://doi.org/10.1103/PhysRevD.108.032002}{\emph{Phys. Rev. D} {\bfseries 108} (2023) 032002} [\href{https://arxiv.org/abs/2212.09153}{{\ttfamily 2212.09153}}].

\bibitem{Bell:2014zya}
G.~Bell and T.~Huber, \emph{{Master integrals for the two-loop penguin contribution in non-leptonic B-decays}}, \href{https://doi.org/10.1007/JHEP12(2014)129}{\emph{JHEP} {\bfseries 12} (2014) 129} [\href{https://arxiv.org/abs/1410.2804}{{\ttfamily 1410.2804}}].

\bibitem{BaBar:2012mrf}
{\scshape BaBar} collaboration, \emph{{Measurement of Branching Fractions and Rate Asymmetries in the Rare Decays $B \to K^{(*)} l^+ l^-$}}, \href{https://doi.org/10.1103/PhysRevD.86.032012}{\emph{Phys. Rev. D} {\bfseries 86} (2012) 032012} [\href{https://arxiv.org/abs/1204.3933}{{\ttfamily 1204.3933}}].

\bibitem{LHCb:2014cxe}
{\scshape LHCb} collaboration, \emph{{Differential branching fractions and isospin asymmetries of $B \to K^{(*)} \mu^+ \mu^-$ decays}}, \href{https://doi.org/10.1007/JHEP06(2014)133}{\emph{JHEP} {\bfseries 06} (2014) 133} [\href{https://arxiv.org/abs/1403.8044}{{\ttfamily 1403.8044}}].

\bibitem{CMS:2024syx}
{\scshape CMS} collaboration, \emph{{Test of lepton flavor universality in B$^{\pm}$$\to$ K$^{\pm}\mu^+\mu^-$ and B$^{\pm}$$\to$ K$^{\pm}$e$^+$e$^-$ decays in proton-proton collisions at $\sqrt{s}$ = 13 TeV}}, \href{https://doi.org/10.1088/1361-6633/ad4e65}{\emph{Rept. Prog. Phys.} {\bfseries 87} (2024) 077802} [\href{https://arxiv.org/abs/2401.07090}{{\ttfamily 2401.07090}}].

\bibitem{LHCb:2020gog}
{\scshape LHCb} collaboration, \emph{{Angular Analysis of the $B^{+}\rightarrow K^{\ast+}\mu^{+}\mu^{-}$ Decay}}, \href{https://doi.org/10.1103/PhysRevLett.126.161802}{\emph{Phys. Rev. Lett.} {\bfseries 126} (2021) 161802} [\href{https://arxiv.org/abs/2012.13241}{{\ttfamily 2012.13241}}].

\bibitem{LHCb:2020lmf}
{\scshape LHCb} collaboration, \emph{{Measurement of $CP$-Averaged Observables in the $B^{0}\rightarrow K^{*0}\mu^{+}\mu^{-}$ Decay}}, \href{https://doi.org/10.1103/PhysRevLett.125.011802}{\emph{Phys. Rev. Lett.} {\bfseries 125} (2020) 011802} [\href{https://arxiv.org/abs/2003.04831}{{\ttfamily 2003.04831}}].

\bibitem{CMS:2017rzx}
{\scshape CMS} collaboration, \emph{{Measurement of angular parameters from the decay $\mathrm{B}^0 \to \mathrm{K}^{*0} \mu^+ \mu^-$ in proton-proton collisions at $\sqrt{s} = $ 8 TeV}}, \href{https://doi.org/10.1016/j.physletb.2018.04.030}{\emph{Phys. Lett. B} {\bfseries 781} (2018) 517} [\href{https://arxiv.org/abs/1710.02846}{{\ttfamily 1710.02846}}].

\bibitem{Belle:2016fev}
{\scshape Belle} collaboration, \emph{{Lepton-Flavor-Dependent Angular Analysis of $B\to K^\ast \ell^+\ell^-$}}, \href{https://doi.org/10.1103/PhysRevLett.118.111801}{\emph{Phys. Rev. Lett.} {\bfseries 118} (2017) 111801} [\href{https://arxiv.org/abs/1612.05014}{{\ttfamily 1612.05014}}].

\bibitem{ATLAS:2018gqc}
{\scshape ATLAS} collaboration, \emph{{Angular analysis of $B^0_d \rightarrow K^{*}\mu^+\mu^-$ decays in $pp$ collisions at $\sqrt{s}= 8$ TeV with the ATLAS detector}}, \href{https://doi.org/10.1007/JHEP10(2018)047}{\emph{JHEP} {\bfseries 10} (2018) 047} [\href{https://arxiv.org/abs/1805.04000}{{\ttfamily 1805.04000}}].

\bibitem{CMS-PAS-BPH-21-002}
{\scshape CMS} collaboration, \emph{{Angular analysis of the $B^0 \to K^{*0}(892) \mu^+ \mu^-$ decay at $\sqrt{s}$ = 13 TeV}},  Tech. Rep. CERN, Geneva (2024).

\bibitem{Buchalla:1995vs}
G.~Buchalla, A.J.~Buras and M.E.~Lautenbacher, \emph{{Weak decays beyond leading logarithms}}, \href{https://doi.org/10.1103/RevModPhys.68.1125}{\emph{Rev. Mod. Phys.} {\bfseries 68} (1996) 1125} [\href{https://arxiv.org/abs/hep-ph/9512380}{{\ttfamily hep-ph/9512380}}].

\bibitem{Bobeth:1999mk}
C.~Bobeth, M.~Misiak and J.~Urban, \emph{{Photonic penguins at two loops and $m_t$ dependence of $BR[B \to X_s l^+ l^-]$}}, \href{https://doi.org/10.1016/S0550-3213(00)00007-9}{\emph{Nucl. Phys. B} {\bfseries 574} (2000) 291} [\href{https://arxiv.org/abs/hep-ph/9910220}{{\ttfamily hep-ph/9910220}}].

\bibitem{Bobeth:2001jm}
C.~Bobeth, A.J.~Buras, F.~Kruger and J.~Urban, \emph{{QCD corrections to $\bar{B} \to X_{d,s} \nu \bar{\nu}$, $\bar{B}_{d,s} \to \ell^{+} \ell^{-}$, $K \to \pi \nu \bar{\nu}$ and $K_{L} \to \mu^{+} \mu^{-}$ in the MSSM}}, \href{https://doi.org/10.1016/S0550-3213(02)00141-4}{\emph{Nucl. Phys. B} {\bfseries 630} (2002) 87} [\href{https://arxiv.org/abs/hep-ph/0112305}{{\ttfamily hep-ph/0112305}}].

\bibitem{Chetyrkin:1997gb}
K.G.~Chetyrkin, M.~Misiak and M.~Munz, \emph{{$|\Delta F| = 1$ nonleptonic effective Hamiltonian in a simpler scheme}}, \href{https://doi.org/10.1016/S0550-3213(98)00131-X}{\emph{Nucl. Phys. B} {\bfseries 520} (1998) 279} [\href{https://arxiv.org/abs/hep-ph/9711280}{{\ttfamily hep-ph/9711280}}].

\bibitem{Khodjamirian:2010vf}
A.~Khodjamirian, T.~Mannel, A.A.~Pivovarov and Y.M.~Wang, \emph{{Charm-loop effect in $B \to K^{(*)} \ell^{+} \ell^{-}$ and $B\to K^*\gamma$}}, \href{https://doi.org/10.1007/JHEP09(2010)089}{\emph{JHEP} {\bfseries 09} (2010) 089} [\href{https://arxiv.org/abs/1006.4945}{{\ttfamily 1006.4945}}].

\bibitem{Gubernari:2020eft}
N.~Gubernari, D.~van Dyk and J.~Virto, \emph{{Non-local matrix elements in $B_{(s)}\to \{K^{(*)},\phi\}\ell^+\ell^-$}}, \href{https://doi.org/10.1007/JHEP02(2021)088}{\emph{JHEP} {\bfseries 02} (2021) 088} [\href{https://arxiv.org/abs/2011.09813}{{\ttfamily 2011.09813}}].

\bibitem{Gubernari:2022hxn}
N.~Gubernari, M.~Reboud, D.~van Dyk and J.~Virto, \emph{{Improved theory predictions and global analysis of exclusive $b \to s\mu^+\mu^-$ processes}}, \href{https://doi.org/10.1007/JHEP09(2022)133}{\emph{JHEP} {\bfseries 09} (2022) 133} [\href{https://arxiv.org/abs/2206.03797}{{\ttfamily 2206.03797}}].

\bibitem{Isidori:2020acz}
G.~Isidori, S.~Nabeebaccus and R.~Zwicky, \emph{{QED corrections in $ \bar{B} \to \bar{K}{\mathrm{\ell}}^{+}{\mathrm{\ell}}^{-} $ at the double-differential level}}, \href{https://doi.org/10.1007/JHEP12(2020)104}{\emph{JHEP} {\bfseries 12} (2020) 104} [\href{https://arxiv.org/abs/2009.00929}{{\ttfamily 2009.00929}}].

\bibitem{Isidori:2022bzw}
G.~Isidori, D.~Lancierini, S.~Nabeebaccus and R.~Zwicky, \emph{{QED in $ \bar{B} $\textrightarrow{}$ \bar{K} $\ensuremath{\ell}$^{+}$\ensuremath{\ell}$^{-}$ LFU ratios: theory versus experiment, a Monte Carlo study}}, \href{https://doi.org/10.1007/JHEP10(2022)146}{\emph{JHEP} {\bfseries 10} (2022) 146} [\href{https://arxiv.org/abs/2205.08635}{{\ttfamily 2205.08635}}].

\bibitem{Choudhury:2023uhw}
D.~Choudhury, D.~Das and J.~Das, \emph{{Soft photon corrections in $B \to K^{(\ast)} \ell^+ \ell^-$ and $\Lambda_b \to \Lambda^{(\ast)} \ell^+ \ell^-$ decays}},  \href{https://arxiv.org/abs/2307.07578}{{\ttfamily 2307.07578}}.

\bibitem{Golonka:2005pn}
P.~Golonka and Z.~Was, \emph{{PHOTOS Monte Carlo: A Precision tool for QED corrections in $Z$ and $W$ decays}}, \href{https://doi.org/10.1140/epjc/s2005-02396-4}{\emph{Eur. Phys. J. C} {\bfseries 45} (2006) 97} [\href{https://arxiv.org/abs/hep-ph/0506026}{{\ttfamily hep-ph/0506026}}].

\bibitem{Mahmoudi:2008tp}
F.~Mahmoudi, \emph{{SuperIso v2.3: A Program for calculating flavor physics observables in Supersymmetry}}, \href{https://doi.org/10.1016/j.cpc.2009.02.017}{\emph{Comput. Phys. Commun.} {\bfseries 180} (2009) 1579} [\href{https://arxiv.org/abs/0808.3144}{{\ttfamily 0808.3144}}].

\bibitem{Becirevic:2012fy}
D.~Becirevic, N.~Kosnik, F.~Mescia and E.~Schneider, \emph{{Complementarity of the constraints on New Physics from $B_s \to \mu^+ \mu^-$ and from $B \to K l^+l^-$ decays}}, \href{https://doi.org/10.1103/PhysRevD.86.034034}{\emph{Phys. Rev. D} {\bfseries 86} (2012) 034034} [\href{https://arxiv.org/abs/1205.5811}{{\ttfamily 1205.5811}}].

\bibitem{Bobeth:2007dw}
C.~Bobeth, G.~Hiller and G.~Piranishvili, \emph{{Angular distributions of $\bar{B} \to \bar{K} \ell^+\ell^-$ decays}}, \href{https://doi.org/10.1088/1126-6708/2007/12/040}{\emph{JHEP} {\bfseries 12} (2007) 040} [\href{https://arxiv.org/abs/0709.4174}{{\ttfamily 0709.4174}}].

\bibitem{Kim:2000dq}
C.S.~Kim, Y.G.~Kim, C.-D.~Lu and T.~Morozumi, \emph{{Azimuthal angle distribution in $B \to K* (\to K \pi) l^+ l^-$ at low invariant $m(l^+ l^-)$ region}}, \href{https://doi.org/10.1103/PhysRevD.62.034013}{\emph{Phys. Rev. D} {\bfseries 62} (2000) 034013} [\href{https://arxiv.org/abs/hep-ph/0001151}{{\ttfamily hep-ph/0001151}}].

\bibitem{Altmannshofer:2008dz}
W.~Altmannshofer, P.~Ball, A.~Bharucha, A.J.~Buras, D.M.~Straub and M.~Wick, \emph{{Symmetries and Asymmetries of $B \to K^{*} \mu^{+} \mu^{-}$ Decays in the Standard Model and Beyond}}, \href{https://doi.org/10.1088/1126-6708/2009/01/019}{\emph{JHEP} {\bfseries 01} (2009) 019} [\href{https://arxiv.org/abs/0811.1214}{{\ttfamily 0811.1214}}].

\bibitem{Kruger:1999xa}
F.~Kruger, L.M.~Sehgal, N.~Sinha and R.~Sinha, \emph{{Angular distribution and CP asymmetries in the decays $\bar B \to K^- \pi^+ e^- e^+$ and $\bar B \to \pi^- \pi^+ e^- e^+$}}, \href{https://doi.org/10.1103/PhysRevD.61.114028}{\emph{Phys. Rev. D} {\bfseries 61} (2000) 114028} [\href{https://arxiv.org/abs/hep-ph/9907386}{{\ttfamily hep-ph/9907386}}].

\bibitem{Egede:2008uy}
U.~Egede, T.~Hurth, J.~Matias, M.~Ramon and W.~Reece, \emph{{New observables in the decay mode $\bar B_d \to \bar K^{*0} l^+ l^-$}}, \href{https://doi.org/10.1088/1126-6708/2008/11/032}{\emph{JHEP} {\bfseries 11} (2008) 032} [\href{https://arxiv.org/abs/0807.2589}{{\ttfamily 0807.2589}}].

\bibitem{Bobeth:2008ij}
C.~Bobeth, G.~Hiller and G.~Piranishvili, \emph{{CP Asymmetries in bar $B \to \bar{K}^* (\to \bar{K} \pi) \bar{\ell} \ell$ and Untagged $\bar{B}_s$, $B_s \to \phi (\to K^{+} K^-) \bar{\ell} \ell$ Decays at NLO}}, \href{https://doi.org/10.1088/1126-6708/2008/07/106}{\emph{JHEP} {\bfseries 07} (2008) 106} [\href{https://arxiv.org/abs/0805.2525}{{\ttfamily 0805.2525}}].

\bibitem{Egede:2010zc}
U.~Egede, T.~Hurth, J.~Matias, M.~Ramon and W.~Reece, \emph{{New physics reach of the decay mode $\bar{B} \to \bar{K}^{*0}\ell^+\ell^-$}}, \href{https://doi.org/10.1007/JHEP10(2010)056}{\emph{JHEP} {\bfseries 10} (2010) 056} [\href{https://arxiv.org/abs/1005.0571}{{\ttfamily 1005.0571}}].

\bibitem{Gratrex:2015hna}
J.~Gratrex, M.~Hopfer and R.~Zwicky, \emph{{Generalised helicity formalism, higher moments and the $B \to K_{J_K}(\to K \pi) \bar{\ell}_1 \ell_2$ angular distributions}}, \href{https://doi.org/10.1103/PhysRevD.93.054008}{\emph{Phys. Rev. D} {\bfseries 93} (2016) 054008} [\href{https://arxiv.org/abs/1506.03970}{{\ttfamily 1506.03970}}].

\bibitem{Matias:2012xw}
J.~Matias, F.~Mescia, M.~Ramon and J.~Virto, \emph{{Complete Anatomy of $\bar{B}_d \to \bar{K}^{* 0} (\to K \pi)l^+l^-$ and its angular distribution}}, \href{https://doi.org/10.1007/JHEP04(2012)104}{\emph{JHEP} {\bfseries 04} (2012) 104} [\href{https://arxiv.org/abs/1202.4266}{{\ttfamily 1202.4266}}].

\bibitem{Beaujean:2012uj}
F.~Beaujean, C.~Bobeth, D.~van Dyk and C.~Wacker, \emph{{Bayesian Fit of Exclusive $b \to s \bar\ell\ell$ Decays: The Standard Model Operator Basis}}, \href{https://doi.org/10.1007/JHEP08(2012)030}{\emph{JHEP} {\bfseries 08} (2012) 030} [\href{https://arxiv.org/abs/1205.1838}{{\ttfamily 1205.1838}}].

\bibitem{Feldmann:2002iw}
T.~Feldmann and J.~Matias, \emph{{Forward backward and isospin asymmetry for $B \to K^* l^+ l^-$ decay in the standard model and in supersymmetry}}, \href{https://doi.org/10.1088/1126-6708/2003/01/074}{\emph{JHEP} {\bfseries 01} (2003) 074} [\href{https://arxiv.org/abs/hep-ph/0212158}{{\ttfamily hep-ph/0212158}}].

\bibitem{Bobeth:2010wg}
C.~Bobeth, G.~Hiller and D.~van Dyk, \emph{{The Benefits of $\bar{B} \to \bar{K}^* l^+ l^-$ Decays at Low Recoil}}, \href{https://doi.org/10.1007/JHEP07(2010)098}{\emph{JHEP} {\bfseries 07} (2010) 098} [\href{https://arxiv.org/abs/1006.5013}{{\ttfamily 1006.5013}}].

\bibitem{Descotes-Genon:2012isb}
S.~Descotes-Genon, J.~Matias, M.~Ramon and J.~Virto, \emph{{Implications from clean observables for the binned analysis of $B \to K^*\mu^+\mu^-$ at large recoil}}, \href{https://doi.org/10.1007/JHEP01(2013)048}{\emph{JHEP} {\bfseries 01} (2013) 048} [\href{https://arxiv.org/abs/1207.2753}{{\ttfamily 1207.2753}}].

\bibitem{Beneke:2000wa}
M.~Beneke and T.~Feldmann, \emph{{Symmetry breaking corrections to heavy to light B meson form-factors at large recoil}}, \href{https://doi.org/10.1016/S0550-3213(00)00585-X}{\emph{Nucl. Phys. B} {\bfseries 592} (2001) 3} [\href{https://arxiv.org/abs/hep-ph/0008255}{{\ttfamily hep-ph/0008255}}].

\bibitem{Bailey:2015dka}
J.A.~Bailey et~al., \emph{{$B\to Kl^+l^-$ Decay Form Factors from Three-Flavor Lattice QCD}}, \href{https://doi.org/10.1103/PhysRevD.93.025026}{\emph{Phys. Rev. D} {\bfseries 93} (2016) 025026} [\href{https://arxiv.org/abs/1509.06235}{{\ttfamily 1509.06235}}].

\bibitem{Parrott:2022rgu}
{\scshape (HPQCD collaboration)\textsection{}, HPQCD} collaboration, \emph{{B\textrightarrow{}K and D\textrightarrow{}K form factors from fully relativistic lattice QCD}}, \href{https://doi.org/10.1103/PhysRevD.107.014510}{\emph{Phys. Rev. D} {\bfseries 107} (2023) 014510} [\href{https://arxiv.org/abs/2207.12468}{{\ttfamily 2207.12468}}].

\bibitem{Boyd:1994tt}
C.G.~Boyd, B.~Grinstein and R.F.~Lebed, \emph{{Constraints on form-factors for exclusive semileptonic heavy to light meson decays}}, \href{https://doi.org/10.1103/PhysRevLett.74.4603}{\emph{Phys. Rev. Lett.} {\bfseries 74} (1995) 4603} [\href{https://arxiv.org/abs/hep-ph/9412324}{{\ttfamily hep-ph/9412324}}].

\bibitem{Bouchard:2013eph}
{\scshape HPQCD} collaboration, \emph{{Rare decay $B \to K \ell^+ \ell^-$ form factors from lattice QCD}}, \href{https://doi.org/10.1103/PhysRevD.88.054509}{\emph{Phys. Rev. D} {\bfseries 88} (2013) 054509} [\href{https://arxiv.org/abs/1306.2384}{{\ttfamily 1306.2384}}].

\bibitem{Bourrely:2008za}
C.~Bourrely, I.~Caprini and L.~Lellouch, \emph{{Model-independent description of B $ \to $ pi l nu decays and a determination of $\vert$V(ub)$\vert$}}, \href{https://doi.org/10.1103/PhysRevD.82.099902}{\emph{Phys. Rev. D} {\bfseries 79} (2009) 013008} [\href{https://arxiv.org/abs/0807.2722}{{\ttfamily 0807.2722}}].

\bibitem{FlavourLatticeAveragingGroupFLAG:2021npn}
{\scshape Flavour Lattice Averaging Group (FLAG)} collaboration, \emph{{FLAG Review 2021}}, \href{https://doi.org/10.1140/epjc/s10052-022-10536-1}{\emph{Eur. Phys. J. C} {\bfseries 82} (2022) 869} [\href{https://arxiv.org/abs/2111.09849}{{\ttfamily 2111.09849}}].

\bibitem{Horgan:2013hoa}
R.R.~Horgan, Z.~Liu, S.~Meinel and M.~Wingate, \emph{{Lattice QCD calculation of form factors describing the rare decays $B \to K^* \ell^+ \ell^-$ and $B_s \to \phi \ell^+ \ell^-$}}, \href{https://doi.org/10.1103/PhysRevD.89.094501}{\emph{Phys. Rev. D} {\bfseries 89} (2014) 094501} [\href{https://arxiv.org/abs/1310.3722}{{\ttfamily 1310.3722}}].

\bibitem{Horgan:2015vla}
R.R.~Horgan, Z.~Liu, S.~Meinel and M.~Wingate, \emph{{Rare $B$ decays using lattice QCD form factors}}, \href{https://doi.org/10.22323/1.214.0372}{\emph{PoS} {\bfseries LATTICE2014} (2015) 372} [\href{https://arxiv.org/abs/1501.00367}{{\ttfamily 1501.00367}}].

\bibitem{Braun:1999dp}
V.M.~Braun, \emph{{QCD sum rules for heavy flavors}}, \href{https://doi.org/10.22323/1.003.0006}{\emph{PoS} {\bfseries hf8} (1999) 006} [\href{https://arxiv.org/abs/hep-ph/9911206}{{\ttfamily hep-ph/9911206}}].

\bibitem{Ball:2004rg}
P.~Ball and R.~Zwicky, \emph{{$B_{d,s} \to \rho, \omega, K^*, \phi$ decay form-factors from light-cone sum rules revisited}}, \href{https://doi.org/10.1103/PhysRevD.71.014029}{\emph{Phys. Rev. D} {\bfseries 71} (2005) 014029} [\href{https://arxiv.org/abs/hep-ph/0412079}{{\ttfamily hep-ph/0412079}}].

\bibitem{Ball:2004ye}
P.~Ball and R.~Zwicky, \emph{{New results on $B \to \pi, K, \eta$ decay formfactors from light-cone sum rules}}, \href{https://doi.org/10.1103/PhysRevD.71.014015}{\emph{Phys. Rev. D} {\bfseries 71} (2005) 014015} [\href{https://arxiv.org/abs/hep-ph/0406232}{{\ttfamily hep-ph/0406232}}].

\bibitem{Duplancic:2008tk}
G.~Duplancic and B.~Melic, \emph{{B, B(s) $ \to $ K form factors: An Update of light-cone sum rule results}}, \href{https://doi.org/10.1103/PhysRevD.78.054015}{\emph{Phys. Rev. D} {\bfseries 78} (2008) 054015} [\href{https://arxiv.org/abs/0805.4170}{{\ttfamily 0805.4170}}].

\bibitem{Bharucha:2015bzk}
A.~Bharucha, D.M.~Straub and R.~Zwicky, \emph{{$B\to V\ell^+\ell^-$ in the Standard Model from light-cone sum rules}}, \href{https://doi.org/10.1007/JHEP08(2016)098}{\emph{JHEP} {\bfseries 08} (2016) 098} [\href{https://arxiv.org/abs/1503.05534}{{\ttfamily 1503.05534}}].

\bibitem{Khodjamirian:2017fxg}
A.~Khodjamirian and A.V.~Rusov, \emph{{$B_{s}\to K \ell \nu_\ell$ and $B_{(s)} \to \pi (K) \ell^+\ell^-$ decays at large recoil and CKM matrix elements}}, \href{https://doi.org/10.1007/JHEP08(2017)112}{\emph{JHEP} {\bfseries 08} (2017) 112} [\href{https://arxiv.org/abs/1703.04765}{{\ttfamily 1703.04765}}].

\bibitem{Khodjamirian:2006st}
A.~Khodjamirian, T.~Mannel and N.~Offen, \emph{{Form-factors from light-cone sum rules with B-meson distribution amplitudes}}, \href{https://doi.org/10.1103/PhysRevD.75.054013}{\emph{Phys. Rev. D} {\bfseries 75} (2007) 054013} [\href{https://arxiv.org/abs/hep-ph/0611193}{{\ttfamily hep-ph/0611193}}].

\bibitem{Lu:2018cfc}
C.-D.~L\"u, Y.-L.~Shen, Y.-M.~Wang and Y.-B.~Wei, \emph{{QCD calculations of $B \to \pi, K$ form factors with higher-twist corrections}}, \href{https://doi.org/10.1007/JHEP01(2019)024}{\emph{JHEP} {\bfseries 01} (2019) 024} [\href{https://arxiv.org/abs/1810.00819}{{\ttfamily 1810.00819}}].

\bibitem{Cui:2022zwm}
B.-Y.~Cui, Y.-K.~Huang, Y.-L.~Shen, C.~Wang and Y.-M.~Wang, \emph{{Precision calculations of B$_{d,s}$ \textrightarrow{} \ensuremath{\pi}, K decay form factors in soft-collinear effective theory}}, \href{https://doi.org/10.1007/JHEP03(2023)140}{\emph{JHEP} {\bfseries 03} (2023) 140} [\href{https://arxiv.org/abs/2212.11624}{{\ttfamily 2212.11624}}].

\bibitem{Gubernari:2018wyi}
N.~Gubernari, A.~Kokulu and D.~van Dyk, \emph{{$B\to P$ and $B\to V$ Form Factors from $B$-Meson Light-Cone Sum Rules beyond Leading Twist}}, \href{https://doi.org/10.1007/JHEP01(2019)150}{\emph{JHEP} {\bfseries 01} (2019) 150} [\href{https://arxiv.org/abs/1811.00983}{{\ttfamily 1811.00983}}].

\bibitem{Monceaux:2023byy}
Y.~Monceaux, A.~Carvunis and F.~Mahmoudi, \emph{{LCSR predictions for $b \to s$ hadronic form factors}}, \href{https://doi.org/10.22323/1.445.0060}{\emph{PoS} {\bfseries FPCP2023} (2023) 060}.

\bibitem{Carvunis:2024koh}
A.~Carvunis, F.~Mahmoudi and Y.~Monceaux, \emph{{On the potential of Light-Cone Sum Rules without Quark-Hadron Duality}},  \href{https://arxiv.org/abs/2404.01290}{{\ttfamily 2404.01290}}.

\bibitem{Descotes-Genon:2019bud}
S.~Descotes-Genon, A.~Khodjamirian and J.~Virto, \emph{{Light-cone sum rules for $B\to K\pi$ form factors and applications to rare decays}}, \href{https://doi.org/10.1007/JHEP12(2019)083}{\emph{JHEP} {\bfseries 12} (2019) 083} [\href{https://arxiv.org/abs/1908.02267}{{\ttfamily 1908.02267}}].

\bibitem{Descotes-Genon:2023ukb}
S.~Descotes-Genon, A.~Khodjamirian, J.~Virto and K.K.~Vos, \emph{{Light-Cone Sum Rules for $S$-wave $B\to K\pi$ Form Factors}}, \href{https://doi.org/10.1007/JHEP06(2023)034}{\emph{JHEP} {\bfseries 06} (2023) 034} [\href{https://arxiv.org/abs/2304.02973}{{\ttfamily 2304.02973}}].

\bibitem{Wang:2015vgv}
Y.-M.~Wang and Y.-L.~Shen, \emph{{QCD corrections to B \textrightarrow{} \ensuremath{\pi} form factors from light-cone sum rules}}, \href{https://doi.org/10.1016/j.nuclphysb.2015.07.016}{\emph{Nucl. Phys. B} {\bfseries 898} (2015) 563} [\href{https://arxiv.org/abs/1506.00667}{{\ttfamily 1506.00667}}].

\bibitem{Braun:2003wx}
V.M.~Braun, D.Y.~Ivanov and G.P.~Korchemsky, \emph{{The B meson distribution amplitude in QCD}}, \href{https://doi.org/10.1103/PhysRevD.69.034014}{\emph{Phys. Rev. D} {\bfseries 69} (2004) 034014} [\href{https://arxiv.org/abs/hep-ph/0309330}{{\ttfamily hep-ph/0309330}}].

\bibitem{Khodjamirian:2020hob}
A.~Khodjamirian, R.~Mandal and T.~Mannel, \emph{{Inverse moment of the B$_{s}$-meson distribution amplitude from QCD sum rule}}, \href{https://doi.org/10.1007/JHEP10(2020)043}{\emph{JHEP} {\bfseries 10} (2020) 043} [\href{https://arxiv.org/abs/2008.03935}{{\ttfamily 2008.03935}}].

\bibitem{Nishikawa:2011qk}
T.~Nishikawa and K.~Tanaka, \emph{{QCD Sum Rules for Quark-Gluon Three-Body Components in the B Meson}}, \href{https://doi.org/10.1016/j.nuclphysb.2013.12.007}{\emph{Nucl. Phys. B} {\bfseries 879} (2014) 110} [\href{https://arxiv.org/abs/1109.6786}{{\ttfamily 1109.6786}}].

\bibitem{Rahimi:2020zzo}
M.~Rahimi and M.~Wald, \emph{{QCD sum rules for parameters of the B-meson distribution amplitudes}}, \href{https://doi.org/10.1103/PhysRevD.104.016027}{\emph{Phys. Rev. D} {\bfseries 104} (2021) 016027} [\href{https://arxiv.org/abs/2012.12165}{{\ttfamily 2012.12165}}].

\bibitem{Colangelo:2000dp}
P.~Colangelo and A.~Khodjamirian, \emph{{QCD sum rules, a modern perspective}},  in \emph{{At the frontier of particle physics. Handbook of QCD. Vol. 1-3}}, M.~Shifman and B.~Ioffe, eds., (Singapore), pp.~1495--1576, World Scientific (2000) [\href{https://arxiv.org/abs/hep-ph/0010175}{{\ttfamily hep-ph/0010175}}].

\bibitem{Caprini:1997mu}
I.~Caprini, L.~Lellouch and M.~Neubert, \emph{{Dispersive bounds on the shape of anti-B $\to$ D(*) lepton anti-neutrino form-factors}}, \href{https://doi.org/10.1016/S0550-3213(98)00350-2}{\emph{Nucl. Phys. B} {\bfseries 530} (1998) 153} [\href{https://arxiv.org/abs/hep-ph/9712417}{{\ttfamily hep-ph/9712417}}].

\bibitem{deRafael:1992tu}
E.~de~Rafael and J.~Taron, \emph{{Constraints on heavy meson form-factors}}, \href{https://doi.org/10.1016/0370-2693(92)90504-W}{\emph{Phys. Lett. B} {\bfseries 282} (1992) 215}.

\bibitem{deRafael:1993ib}
E.~de~Rafael and J.~Taron, \emph{{Analyticity properties and unitarity constraints of heavy meson form-factors}}, \href{https://doi.org/10.1103/PhysRevD.50.373}{\emph{Phys. Rev. D} {\bfseries 50} (1994) 373} [\href{https://arxiv.org/abs/hep-ph/9306214}{{\ttfamily hep-ph/9306214}}].

\bibitem{Boyd:1995cf}
C.G.~Boyd, B.~Grinstein and R.F.~Lebed, \emph{{Model independent extraction of $\vert$V(cb)$\vert$ using dispersion relations}}, \href{https://doi.org/10.1016/0370-2693(95)00480-9}{\emph{Phys. Lett. B} {\bfseries 353} (1995) 306} [\href{https://arxiv.org/abs/hep-ph/9504235}{{\ttfamily hep-ph/9504235}}].

\bibitem{Gubernari:2023puw}
N.~Gubernari, M.~Reboud, D.~van Dyk and J.~Virto, \emph{{Dispersive analysis of B \textrightarrow{} K$^{(*)}$ and B$_{s}$\textrightarrow{} \ensuremath{\phi} form factors}}, \href{https://doi.org/10.1007/JHEP12(2023)153}{\emph{JHEP} {\bfseries 12} (2023) 153} [\href{https://arxiv.org/abs/2305.06301}{{\ttfamily 2305.06301}}].

\bibitem{Beneke:2001at}
M.~Beneke, T.~Feldmann and D.~Seidel, \emph{{Systematic approach to exclusive $B \to V l^+ l^-$, $V \gamma$ decays}}, \href{https://doi.org/10.1016/S0550-3213(01)00366-2}{\emph{Nucl. Phys. B} {\bfseries 612} (2001) 25} [\href{https://arxiv.org/abs/hep-ph/0106067}{{\ttfamily hep-ph/0106067}}].

\bibitem{Beneke:2004dp}
M.~Beneke, T.~Feldmann and D.~Seidel, \emph{{Exclusive radiative and electroweak $b \to d$ and $b \to s$ penguin decays at NLO}}, \href{https://doi.org/10.1140/epjc/s2005-02181-5}{\emph{Eur. Phys. J. C} {\bfseries 41} (2005) 173} [\href{https://arxiv.org/abs/hep-ph/0412400}{{\ttfamily hep-ph/0412400}}].

\bibitem{Hurth:2013ssa}
T.~Hurth and F.~Mahmoudi, \emph{{On the LHCb anomaly in $B \to K^*\ell^+\ell^-$}}, \href{https://doi.org/10.1007/JHEP04(2014)097}{\emph{JHEP} {\bfseries 04} (2014) 097} [\href{https://arxiv.org/abs/1312.5267}{{\ttfamily 1312.5267}}].

\bibitem{Lyon:2014hpa}
J.~Lyon and R.~Zwicky, \emph{{Resonances gone topsy turvy - the charm of QCD or new physics in $b \to s \ell^+ \ell^-$?}},  \href{https://arxiv.org/abs/1406.0566}{{\ttfamily 1406.0566}}.

\bibitem{Brass:2016efg}
S.~Bra\ss{}, G.~Hiller and I.~Nisandzic, \emph{{Zooming in on $B\rightarrow K^*\ell \ell $ decays at low recoil}}, \href{https://doi.org/10.1140/epjc/s10052-016-4576-9}{\emph{Eur. Phys. J. C} {\bfseries 77} (2017) 16} [\href{https://arxiv.org/abs/1606.00775}{{\ttfamily 1606.00775}}].

\bibitem{Blake:2017fyh}
T.~Blake, U.~Egede, P.~Owen, K.A.~Petridis and G.~Pomery, \emph{{An empirical model to determine the hadronic resonance contributions to $\bar{B}{} ^0 \!\rightarrow \bar{K}{} ^{*0} \mu ^+ \mu ^- $ transitions}}, \href{https://doi.org/10.1140/epjc/s10052-018-5937-3}{\emph{Eur. Phys. J. C} {\bfseries 78} (2018) 453} [\href{https://arxiv.org/abs/1709.03921}{{\ttfamily 1709.03921}}].

\bibitem{Cornella:2020aoq}
C.~Cornella, G.~Isidori, M.~K\"onig, S.~Liechti, P.~Owen and N.~Serra, \emph{{Hunting for $B^+\rightarrow K^+ \tau ^+\tau ^-$ imprints on the $B^+ \rightarrow K^+ \mu ^+\mu ^-$ dimuon spectrum}}, \href{https://doi.org/10.1140/epjc/s10052-020-08674-5}{\emph{Eur. Phys. J. C} {\bfseries 80} (2020) 1095} [\href{https://arxiv.org/abs/2001.04470}{{\ttfamily 2001.04470}}].

\bibitem{Bordone:2024hui}
M.~Bordone, G.~isidori, S.~M\"achler and A.~Tinari, \emph{{Short- vs. long-distance physics in $B\to K^{(*)} \ell^+\ell^-$: a data-driven analysis}},  \href{https://arxiv.org/abs/2401.18007}{{\ttfamily 2401.18007}}.

\bibitem{Asatrian:2019kbk}
H.M.~Asatrian, C.~Greub and J.~Virto, \emph{{Exact NLO matching and analyticity in $b\to s\ell\ell$}}, \href{https://doi.org/10.1007/JHEP04(2020)012}{\emph{JHEP} {\bfseries 04} (2020) 012} [\href{https://arxiv.org/abs/1912.09099}{{\ttfamily 1912.09099}}].

\bibitem{Asatryan:2001zw}
H.H.~Asatryan, H.M.~Asatrian, C.~Greub and M.~Walker, \emph{{Calculation of two loop virtual corrections to $b \to s l^+ l^-$ in the standard model}}, \href{https://doi.org/10.1103/PhysRevD.65.074004}{\emph{Phys. Rev. D} {\bfseries 65} (2002) 074004} [\href{https://arxiv.org/abs/hep-ph/0109140}{{\ttfamily hep-ph/0109140}}].

\bibitem{Greub:2008cy}
C.~Greub, V.~Pilipp and C.~Schupbach, \emph{{Analytic calculation of two-loop QCD corrections to $b \to sl^+ l^-$ in the high $q^2$ region}}, \href{https://doi.org/10.1088/1126-6708/2008/12/040}{\emph{JHEP} {\bfseries 12} (2008) 040} [\href{https://arxiv.org/abs/0810.4077}{{\ttfamily 0810.4077}}].

\bibitem{Ghinculov:2003qd}
A.~Ghinculov, T.~Hurth, G.~Isidori and Y.P.~Yao, \emph{{The Rare decay $B \to X_s l^+ l^-$ to NNLL precision for arbitrary dilepton invariant mass}}, \href{https://doi.org/10.1016/j.nuclphysb.2004.02.028}{\emph{Nucl. Phys. B} {\bfseries 685} (2004) 351} [\href{https://arxiv.org/abs/hep-ph/0312128}{{\ttfamily hep-ph/0312128}}].

\bibitem{deBoer:2017way}
S.~de~Boer, \emph{{Two loop virtual corrections to $b\rightarrow (d,s)\ell ^+\ell ^-$ and $c\rightarrow u\ell ^+\ell ^-$ for arbitrary momentum transfer}}, \href{https://doi.org/10.1140/epjc/s10052-017-5364-x}{\emph{Eur. Phys. J. C} {\bfseries 77} (2017) 801} [\href{https://arxiv.org/abs/1707.00988}{{\ttfamily 1707.00988}}].

\bibitem{LHCb:2016due}
{\scshape LHCb} collaboration, \emph{{Measurement of the phase difference between short- and long-distance amplitudes in the $B^{+}\to K^{+}\mu^{+}\mu^{-}$ decay}}, \href{https://doi.org/10.1140/epjc/s10052-017-4703-2}{\emph{Eur. Phys. J. C} {\bfseries 77} (2017) 161} [\href{https://arxiv.org/abs/1612.06764}{{\ttfamily 1612.06764}}].

\bibitem{LHCb:2023gel}
{\scshape LHCb} collaboration, \emph{{Determination of short- and long-distance contributions in $B^0 \to K^{*0}\ensuremath{\mu}^+\ensuremath{\mu}^-$ decays}}, \href{https://doi.org/10.1103/PhysRevD.109.052009}{\emph{Phys. Rev. D} {\bfseries 109} (2024) 052009} [\href{https://arxiv.org/abs/2312.09102}{{\ttfamily 2312.09102}}].

\bibitem{LHCb:2024onj}
{\scshape LHCb} collaboration, \emph{{Comprehensive analysis of local and nonlocal amplitudes in the $B^0\rightarrow K^{*0}\mu^+\mu^-$ decay}},  \href{https://arxiv.org/abs/2405.17347}{{\ttfamily 2405.17347}}.

\bibitem{Ciuchini:2022wbq}
M.~Ciuchini, M.~Fedele, E.~Franco, A.~Paul, L.~Silvestrini and M.~Valli, \emph{{Constraints on lepton universality violation from rare B decays}}, \href{https://doi.org/10.1103/PhysRevD.107.055036}{\emph{Phys. Rev. D} {\bfseries 107} (2023) 055036} [\href{https://arxiv.org/abs/2212.10516}{{\ttfamily 2212.10516}}].

\bibitem{Isidori:2024lng}
G.~Isidori, Z.~Polonsky and A.~Tinari, \emph{{An explicit estimate of charm rescattering in $B^0 \to K^0 \bar{\ell} \ell$}},  \href{https://arxiv.org/abs/2405.17551}{{\ttfamily 2405.17551}}.

\bibitem{Misiak:2004ew}
M.~Misiak and M.~Steinhauser, \emph{{Three loop matching of the dipole operators for $b \to s \gamma$ and $b \to s g$}}, \href{https://doi.org/10.1016/j.nuclphysb.2004.02.006}{\emph{Nucl. Phys. B} {\bfseries 683} (2004) 277} [\href{https://arxiv.org/abs/hep-ph/0401041}{{\ttfamily hep-ph/0401041}}].

\bibitem{Gorbahn:2004my}
M.~Gorbahn and U.~Haisch, \emph{{Effective Hamiltonian for non-leptonic $|\Delta F| = 1$ decays at NNLO in QCD}}, \href{https://doi.org/10.1016/j.nuclphysb.2005.01.047}{\emph{Nucl. Phys. B} {\bfseries 713} (2005) 291} [\href{https://arxiv.org/abs/hep-ph/0411071}{{\ttfamily hep-ph/0411071}}].

\bibitem{Bobeth:2003at}
C.~Bobeth, P.~Gambino, M.~Gorbahn and U.~Haisch, \emph{{Complete NNLO QCD analysis of $\bar{B} \to X(s) l^+ l^-$ and higher order electroweak effects}}, \href{https://doi.org/10.1088/1126-6708/2004/04/071}{\emph{JHEP} {\bfseries 04} (2004) 071} [\href{https://arxiv.org/abs/hep-ph/0312090}{{\ttfamily hep-ph/0312090}}].

\bibitem{Huber:2005ig}
T.~Huber, E.~Lunghi, M.~Misiak and D.~Wyler, \emph{{Electromagnetic logarithms in $\bar B \to X_s l^+ l^-$}}, \href{https://doi.org/10.1016/j.nuclphysb.2006.01.037}{\emph{Nucl. Phys. B} {\bfseries 740} (2006) 105} [\href{https://arxiv.org/abs/hep-ph/0512066}{{\ttfamily hep-ph/0512066}}].

\bibitem{ParticleDataGroup:2022pth}
{\scshape Particle Data Group} collaboration, \emph{{Review of Particle Physics}}, \href{https://doi.org/10.1093/ptep/ptac097}{\emph{PTEP} {\bfseries 2022} (2022) 083C01}.

\bibitem{ValeSilva:2024jml}
L.~Vale~Silva, \emph{{2023 update of the extraction of the CKM matrix elements}},  in \emph{{12th International Workshop on the CKM Unitarity Triangle}}, 5, 2024 [\href{https://arxiv.org/abs/2405.08046}{{\ttfamily 2405.08046}}].

\bibitem{UTfit:2022hsi}
{\scshape UTfit} collaboration, \emph{{New UTfit Analysis of the Unitarity Triangle in the Cabibbo-Kobayashi-Maskawa scheme}}, \href{https://doi.org/10.1007/s12210-023-01137-5}{\emph{Rend. Lincei Sci. Fis. Nat.} {\bfseries 34} (2023) 37} [\href{https://arxiv.org/abs/2212.03894}{{\ttfamily 2212.03894}}].

\bibitem{Mahmoudi:2007vz}
F.~Mahmoudi, \emph{{SuperIso: A Program for calculating the isospin asymmetry of $B \to K^* \gamma$ in the MSSM}}, \href{https://doi.org/10.1016/j.cpc.2007.12.006}{\emph{Comput. Phys. Commun.} {\bfseries 178} (2008) 745} [\href{https://arxiv.org/abs/0710.2067}{{\ttfamily 0710.2067}}].

\bibitem{Neshatpour:2021nbn}
S.~Neshatpour and F.~Mahmoudi, \emph{{Flavour Physics with SuperIso}}, \href{https://doi.org/10.22323/1.392.0036}{\emph{PoS} {\bfseries TOOLS2020} (2021) 036} [\href{https://arxiv.org/abs/2105.03428}{{\ttfamily 2105.03428}}].

\bibitem{Neshatpour:2022fak}
S.~Neshatpour and F.~Mahmoudi, \emph{{Flavour Physics Phenomenology with SuperIso}}, \href{https://doi.org/10.22323/1.409.0010}{\emph{PoS} {\bfseries CompTools2021} (2022) 010} [\href{https://arxiv.org/abs/2207.04956}{{\ttfamily 2207.04956}}].

\bibitem{Hurth:2016fbr}
T.~Hurth, F.~Mahmoudi and S.~Neshatpour, \emph{{On the anomalies in the latest LHCb data}}, \href{https://doi.org/10.1016/j.nuclphysb.2016.05.022}{\emph{Nucl. Phys. B} {\bfseries 909} (2016) 737} [\href{https://arxiv.org/abs/1603.00865}{{\ttfamily 1603.00865}}].

\bibitem{Chobanova:2017ghn}
V.G.~Chobanova, T.~Hurth, F.~Mahmoudi, D.~Martinez~Santos and S.~Neshatpour, \emph{{Large hadronic power corrections or new physics in the rare decay $B \to K^{*}\mu^{+}\mu^{-}$?}}, \href{https://doi.org/10.1007/JHEP07(2017)025}{\emph{JHEP} {\bfseries 07} (2017) 025} [\href{https://arxiv.org/abs/1702.02234}{{\ttfamily 1702.02234}}].

\bibitem{Hurth:2017hxg}
T.~Hurth, F.~Mahmoudi, D.~Martinez~Santos and S.~Neshatpour, \emph{{Lepton nonuniversality in exclusive $b{\rightarrow}s{\ell}{\ell}$ decays}}, \href{https://doi.org/10.1103/PhysRevD.96.095034}{\emph{Phys. Rev. D} {\bfseries 96} (2017) 095034} [\href{https://arxiv.org/abs/1705.06274}{{\ttfamily 1705.06274}}].

\bibitem{Arbey:2018ics}
A.~Arbey, T.~Hurth, F.~Mahmoudi and S.~Neshatpour, \emph{{Hadronic and New Physics Contributions to $b \to s$ Transitions}}, \href{https://doi.org/10.1103/PhysRevD.98.095027}{\emph{Phys. Rev. D} {\bfseries 98} (2018) 095027} [\href{https://arxiv.org/abs/1806.02791}{{\ttfamily 1806.02791}}].

\bibitem{Hurth:2020rzx}
T.~Hurth, F.~Mahmoudi and S.~Neshatpour, \emph{{Implications of the new LHCb angular analysis of $B \to K^* \mu^+ \mu^-$ : Hadronic effects or new physics?}}, \href{https://doi.org/10.1103/PhysRevD.102.055001}{\emph{Phys. Rev. D} {\bfseries 102} (2020) 055001} [\href{https://arxiv.org/abs/2006.04213}{{\ttfamily 2006.04213}}].

\bibitem{Hurth:2023jwr}
T.~Hurth, F.~Mahmoudi and S.~Neshatpour, \emph{{$B$ anomalies in the post $R_{K^{(*)}}$ era}}, \href{https://doi.org/10.1103/PhysRevD.108.115037}{\emph{Phys. Rev. D} {\bfseries 108} (2023) 115037} [\href{https://arxiv.org/abs/2310.05585}{{\ttfamily 2310.05585}}].

\end{thebibliography}\endgroup

\end{document}